\documentclass[journal]{IEEEtran}

\usepackage{cite}
\usepackage{amsmath,amssymb,amsfonts}

\usepackage{algorithm}
\usepackage[noend]{algpseudocode}
\algnewcommand\algorithmicinput{\textbf{Input:}}
\algnewcommand\algorithmicoutput{\textbf{Output:}}
\algnewcommand\Input{\item[\algorithmicinput]}%
\algnewcommand\Output{\item[\algorithmicoutput]}%
\algnewcommand{\LineComment}[1]{\State \(\triangleright\) #1}

\newlength{\subcolumnwidth}

\newcommand{\nextsubcolumn}[1][]{%
  \cr\noalign{\hfill}
  \if\relax\detokenize{#1}\relax\else\hsize=#1\setlength{\subcolumnwidth}{\hsize}\fi
}

\usepackage[dvips]{graphicx}
\usepackage{textcomp}

\ifCLASSOPTIONcompsoc
  \usepackage[caption=false,font=normalsize,labelfont=sf,textfont=sf]{subfig}
\else
  \usepackage[caption=false,font=footnotesize]{subfig}
\fi
\usepackage{placeins}
         
\usepackage[font={sf,scriptsize},
            labelfont={bf},
            justification= raggedright, 
            singlelinecheck=false]
            {caption}

\usepackage{array}
\newcolumntype{C}{>{\centering\arraybackslash}p{0.04\textwidth}}
\newcolumntype{L}{>{\raggedright\arraybackslash}p{0.02\textwidth}}

\usepackage[table]{xcolor}
\definecolor{Grey}{gray}{0.93}
\definecolor{blue}{RGB}{0, 47, 167}

\usepackage{xcolor}
\definecolor{colour_carrier}{HTML}{6C8EBF}
\definecolor{colour_noncarrier}{HTML}{82B366}

\usepackage{bm}
\usepackage{enumitem}
\usepackage{mathtools}
\usepackage{csquotes}

\usepackage{soul}

\begin{document}
%
\title{Cyber Vaccine for Deepfake Immunity}

%
%
%

\author{Ching-Chun~Chang, Huy Hong Nguyen, Junichi Yamagishi, Isao Echizen
\thanks{This work was supported in part by the Japan Society for the Promotion of Science (JSPS) under KAKENHI Grants (JP16H06302, JP18H04120, JP20K23355, JP21H04907 and JP21K18023) and the Japan Science and Technology Agency (JST) under CREST Grants (JPMJCR18A6 and JPMJCR20D3).}
\thanks{C.-C. Chang, H. H. Nguyen, J. Yamagishi and I. Echizen are with the National Institute of Informatics, Tokyo, Japan (email: ccchang@nii.ac.jp; nhhuy@nii.ac.jp; jyamagis@nii.ac.jp; iechizen@nii.ac.jp).}
\thanks{Digital Object Identifier xx.xxxx/2023}
}

%
%

\markboth{Journal of \LaTeX\ Class Files,~Vol.~??, No.~?, Month~Year}%
{Chang \MakeLowercase{\textit{et al.}}: Cyber vaccine}
%



\maketitle

\begin{abstract}
Deepfakes pose an evolving threat to cybersecurity, which calls for the development of automated countermeasures. While considerable forensic research has been devoted to the detection and localisation of deepfakes, solutions for reversing fake to real are yet to be developed. In this study, we introduce cyber vaccination for conferring immunity to deepfakes. Analogous to biological vaccination that injects antigens to induce immunity prior to infection by an actual pathogen, cyber vaccination simulates deepfakes and performs adversarial training to build a defensive immune system. Aiming at building up attack-agnostic immunity with limited computational resources, we propose to simulate various deepfakes with one single overpowered attack: face masking. The proposed immune system consists of a vaccinator for inducing immunity and a neutraliser for recovering facial content. Experimental evaluations demonstrate effective immunity to face replacement, face reenactment and various types of corruptions.
\end{abstract}



%
\IEEEpeerreviewmaketitle

\section{Introduction}\label{sec:intro}
\IEEEPARstart{D}{eepfakes}, as an emergent cyber-security threat, leverage artificial intelligence and machine learning to create synthetic media. This technology is a double-edged sword that can facilitate innocent entertainment, but also entails insidious ramifications to economy, politics and society, including but not limited to market manipulation, electoral influence, nonconsensual pornography, defamatory accusation, fabricated evidence and identity fraud. As the saying goes, `seeing is believing', this natural inclination fuels the spread of disinformation posed by deepfakes. The widespread use and rapid advancement of deepfakes present an evolving challenge to develop countermeasures to tell fact from fiction~\cite{10.1145/3371409}.

\begin{figure}[t!] 
\centering
\subfloat[detection]{\includegraphics[width=0.33\columnwidth]{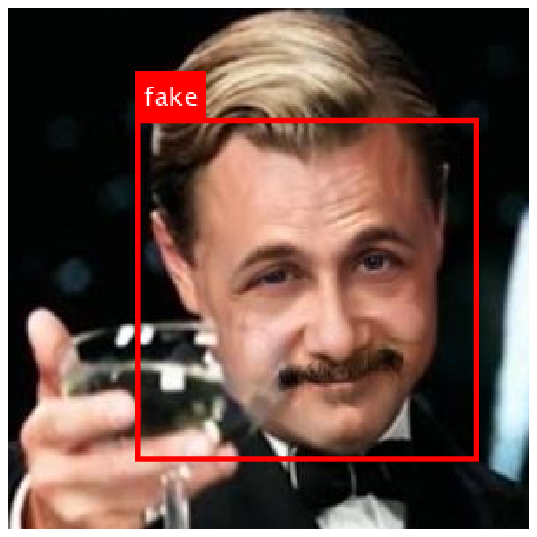}}
\hfil
\subfloat[localisation]{\includegraphics[width=0.33\columnwidth]{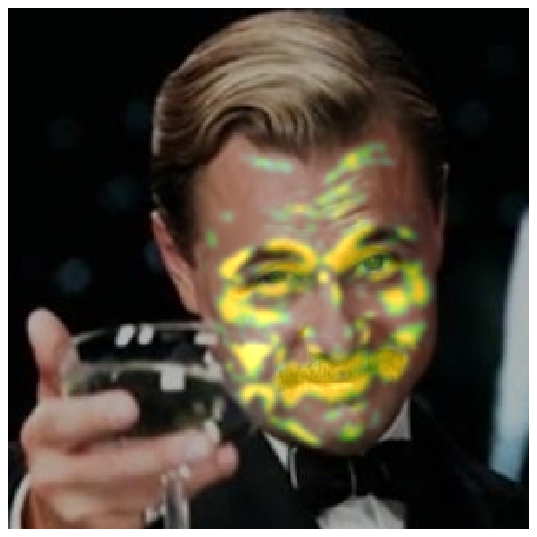}}
\hfil
\subfloat[restoration]{\includegraphics[width=0.33\columnwidth]{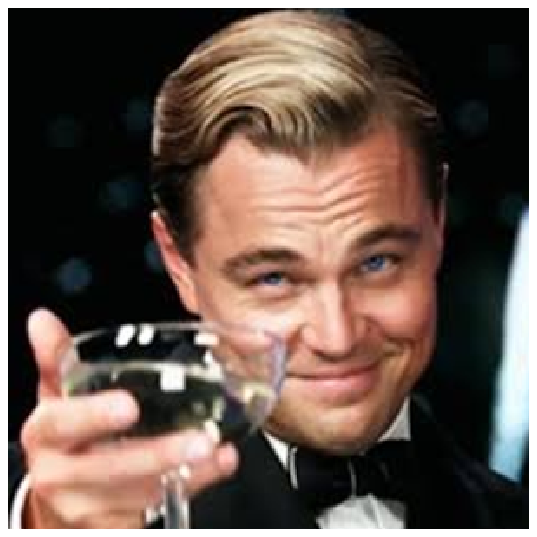}}

\caption{Aims of cyber forensics: detection, localisation, restoration.}
\label{fig:three_tasks}
\end{figure}

While the term deepfakes has been generalised to refer to a broad range of synthetic media nowadays, we focus on a main category that arouses major public concerns\textemdash facial manipulation~\cite{10.1145/3425780}. In general, defence against manipulation centres around three fundamental (step-by-step) tasks: detection, localisation and restoration, as illustrated in Figure~\ref{fig:three_tasks}. Computational forensics has offered fruitful outcomes for detecting whether a given image has been tampered by deepfake algorithms~\cite{8397040, 8630761, 8630787, 8639163, 8682602, 9157215} and further localising the tampered parts~\cite{8237794, 8578214, 8626149, 9185974}. However, solutions for restoring fraudulent content are yet to be developed. Unlike detection and localisation that can rely on passive diagnostics, reliable restoration often requires active safeguard. In digital communications, for instance, error correction codes are used to protect message prior to transmission over a noisy channel. Vaccination, as another analogy, injects antigens to trigger an immune response within the body, thereby inducing protective immunity prior to infection by an actual pathogen.

Adversarial training, similar to the notion of pathogen-specific vaccines, is an intuitive approach towards combating security vulnerabilities~\cite{Madry:2018aa, Tramer:2018aa, 10.5555/3454287.3454814}. This is performed by training models in the presence of simulated adversaries so that the computational models, similar to a biological immune system, learn to defend against possible attacks. While adversarial training can be a potential solution, there are a number of inherent limitations. First of all, preparation and execution of a wide variety of deepfake algorithms can be expensive in terms of (computational) time and resources. On top of this, deepfakes, as a type of cyber viruses, evolve over time and it is virtually impossible to take every possibility into consideration in practice. It is like a vaccine to one specific pathogen, which typically cannot protect against another, except when the two are very similar to each other. Hence, it is important to pursue the capability to defend against various (even unforeseen) attacks, sometimes referred to as attack agnosticism.

In this paper, we study an attack-agnostic method for restoration of deepfakes. The proposed system consists of a vaccinator for inducing immunity, a neutraliser for recovering facial content, a validator for distinguishing between vaccinated and unvaccinated media, and an adversary for simulating deepfakes. Instead of exhausting every possible deepfakes, we consider a single overwhelming adversary model, the masked-face model, in an attempt to build up attack-agnostic capability with limited computational resources.

\begin{figure*}[t]
\centerline{\includegraphics[width=1.85\columnwidth]{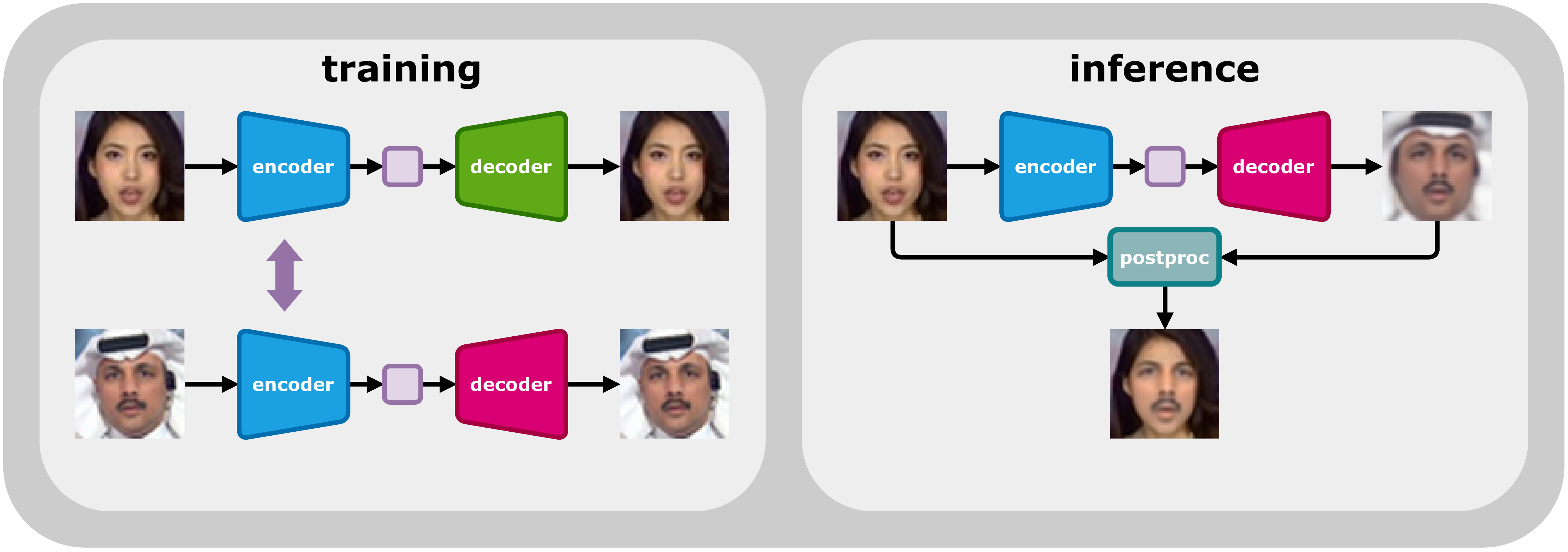}}
\caption{Deepfake autoencoder for face replacement.}
\label{fig:faceswap}
\end{figure*}

\section{Preliminaries}
\label{sec:pre}
The term deepfakes, a portmanteau of deep learning and fake media, is considered originating from an anonymous user named `deepfakes' who posted face-swap videos on a social media platform. Face replacement, as the very first example of deepfakes, creates convincing face-swap videos with an auto-encoder model, which consists of a shared encoder and two decoders for two respective identities (source and target), as illustrated in Figure~\ref{fig:faceswap}. Auto-encoder is a class of neural network models that use an encoder to reduce data dimensionality, or to project data into a compact latent space, and a decoder to reconstruct data from the latent features. This process mimics squeezing information through a bottleneck, thereby retaining useful information for prediction. A heuristic idea of face-swap auto-encoder is that the shared encoder learns to extract features such as varying features such as facial expressions and poses, whereas each decoder learns to use such features along with invariant features about the corresponding identity to reconstruct the video frame. Once trained, the synthetic source frames are generated by passing the target frames through the shared encoder, while reconstructing the video frames with the decoder of the source identity. A face-swap video is then produced by blending the face region in the synthetic source frames with non-face region in the target frames.

Another example of deepfakes is referred to as face reenactment that turns an identity into a puppet. As the name suggests, this technique manipulates a target individual's facial movements such as expression~\cite{10.1145/2816795.2818056, 10.1145/3197517.3201283, 10.1145/3292039}, gaze~\cite{8010348, 8953732, Ganin:2016aa}, pose~\cite{8453880, 8578457, 8578968, 8578974, Neverova:2018aa} and body~\cite{10.1145/3333002, 8658561, 9008556, 9022584, Aberman:2019aa}. Mouth reenactment, also known as lip synchronisation, matches a target’s lip movements with a vocal audio track~\cite{10.1007/978-3-030-58517-4_42, 10.1145/3072959.3073640, Jamaludin:2019aa, Vougioukas:2020aa}. Facial editing or retouching is also a common form of deepfake manipulations that alters the appearance of a target, usually for entertainment~\cite{Brock:2017aa, 8100061, 8718508}.


\section{Methodology}
\label{sec:method}
We begin with a fundamental conceptual model of cyber vaccination and discuss the limitations of a natural solution. We then present the proposed solution as well as the procedures for building an immune system.

\subsection{Deepfake Sampling}
Cyber vaccination can be viewed as a form of communications. A communication system typically consists of an encoder (at the source) and a decoder (at the destination)~\cite{1948_6773024}. The goal is to accurately transmit a message from the source to the destination over a noisy channel with the help of an encoder/decoder pair. As the communication system uses a pair of encoder and decoder for error correction, cyber vaccination system has a pair of vaccinator and neutraliser for manipulation reversal. It is possible to train a pair of neural networks jointly with an attack model in the middle. Random sampling of attack models during the training process may lead to adaptability to various toxic conditions. However, there are a number of challenges that may limit the practicality of this approach.

From an engineering standpoint, it is difficult to construct a universal deepfake toolkit with diverse attack models, which can be incorporated into the training process in real time. Impediments to universal toolkit include, but are not limited to, different input requirements (e.g. photo size, face position, portrait composition, axillary data), different levels of generalisability (i.e. identity-specific or identity-agnostic) and different prep/post-processing procedures. In addition to this, different attack models would cause very different degrees and forms of distortion (e.g. face reenactment and face replacement) and therefore it is arguable whether the training loss can converge within a reasonable time frame. Furthermore, it is a formidable challenge to prove that in-the-lab vaccines can be reliably transferred to protect against in-the-wild virus variants.

\begin{figure*}[t]
\centerline{\includegraphics[width=1.85\columnwidth]{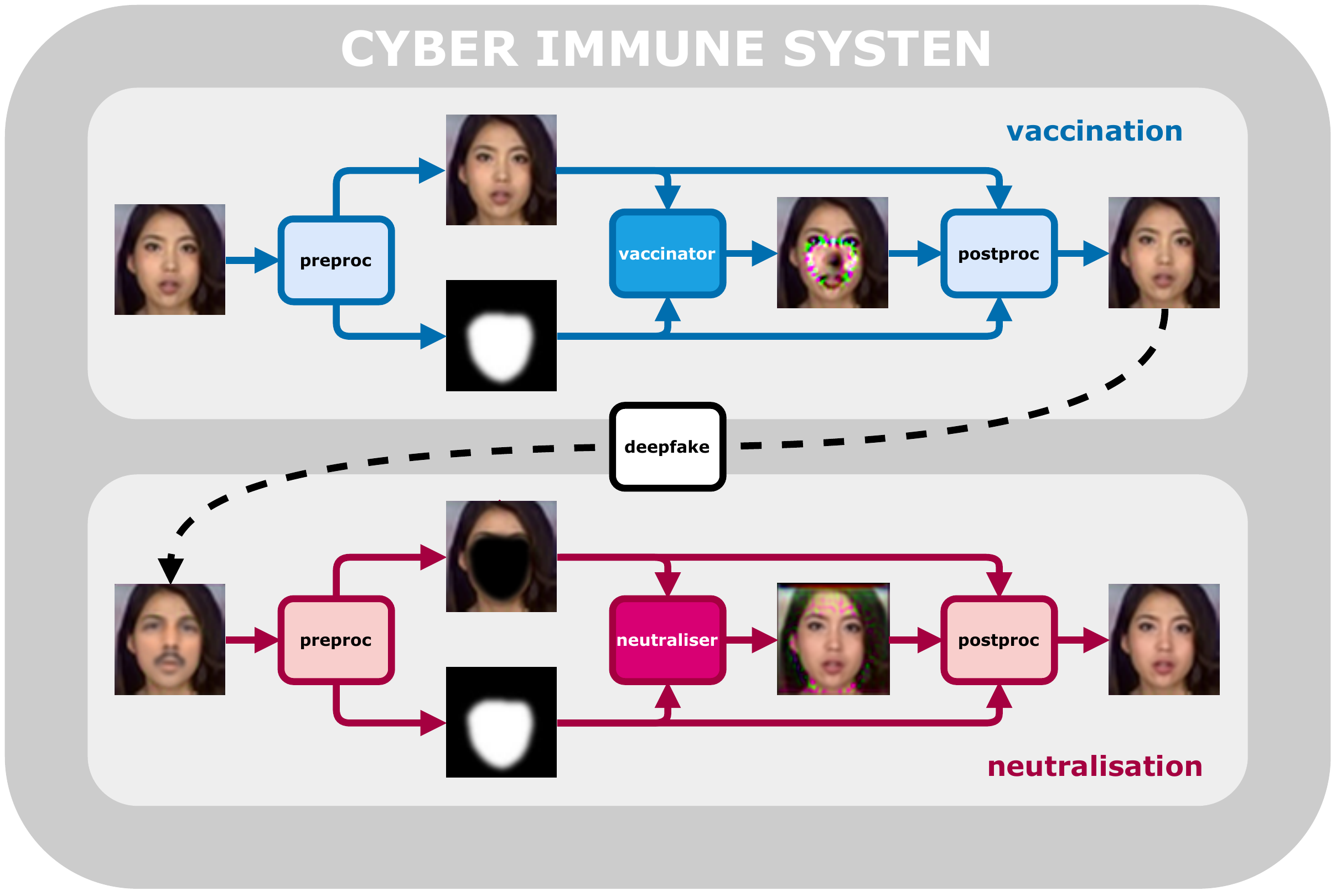}}
\caption{Cyber immune system with a pair of vaccinator and neutraliser for deepfake restoration.}
\label{fig:immune_syst}
\end{figure*}

\subsection{Face Masking}
To overcome these issues, neither restricting on a limited number of deepfake algorithms nor exhausting all possibilities, we attempt to consider one single attack model. It should be identity-agnostic and can be executed in real time. Most importantly, addressing this very fatal attack implies addressing a variety of deepfake attacks. A common factor for nearly all deepfake manipulations is that the face region becomes untrustworthy to a greater or lesser extent. As an extreme case, face masking can be an ideal attack model that satisfies all aforementioned purposes.


If we mask out the face region and attempt to reconstruct it based on the rest context information, this becomes an image inpainting problem~\cite{10.1145/344779.344972, 10.1145/3072959.3073659, 8100107}. Although one could expect a plausible and realistic reconstruction of the missing parts, the original content cannot be recovered with absolute certainty in most of the cases. If we permit imperceptible modifications prior to face masking, we could apply steganographic algorithms to embed the reconstruction information about the face region into the non-face region. Nonetheless, this steganographic solution (also referred to as self-embedding) often requires non-trivial manual adjustments of parameters when putting into practice~\cite{817228, 4694853, 5961626, Qin:2012aa, 6355682, 8259425}. Consider that steganographic capacity is limited under an embedding distortion constraint. For controlling the amount of data to be embedded below the capacity limit, one may apply source coding to compress the data, which involves a trade-off between code efficiency and reconstruction quality. Consider further that digital images may be processed by a wide range of transformations in blurriness, brightness, contrast, saturation, hue, etc. Due to the fragility of steganography, one may apply channel coding to correct errors for ensuring reliable data extraction, which involves a trade-off between code redundancy and correction capability. Moreover, some content-dependent algorithmic parameters may need to be optimised for each individual image. While there are learning-based steganographic methods aiming to embed messages in an automatic and robust manner~\cite{NIPS2017_838e8afb, Zhu:2018aa, Tancik:2020aa}, the message is usually assumed to be a sequence of random binary digits or a secret image and the hiding location cannot be specified. In our context, the message is the face region of the portrait image itself and the hiding location is the non-face region. Synchronisation may also be a problem if the face region detected at the encoder side is inconsistent with that at the decoder side.

\subsection{Cyber Immune System}
Machine learning forges a path to be (mostly) free from manual intervention in parameter configurations. Consider a transmission of a portrait image between a vaccinator and a neutraliser over a face masking channel. Both vaccinator and neutraliser are neural networks and the goal is to preserve the quality vaccinated image while ensuring quality of neutralised image. In addition to this, it is required to be able to distinguish between the vaccinated and unvaccinated objects. Furthermore, robustness against common image processing operations would be an appealing feature in practice. Both vaccination and neutralisation are composed of pre-processing, mid-processing and post-processing steps, as illustrated in Figure~\ref{fig:immune_syst}.

\begin{figure}[t!] 
\centering
\subfloat[unvaccinated samples]{\includegraphics[width=0.98\columnwidth]{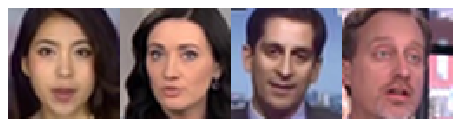}}
\\
\subfloat[vaccinated samples]{\includegraphics[width=0.98\columnwidth]{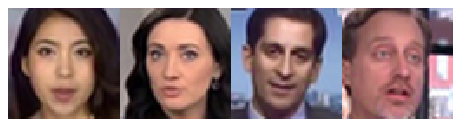}}
\\
\subfloat[residuals]{\includegraphics[width=0.98\columnwidth]{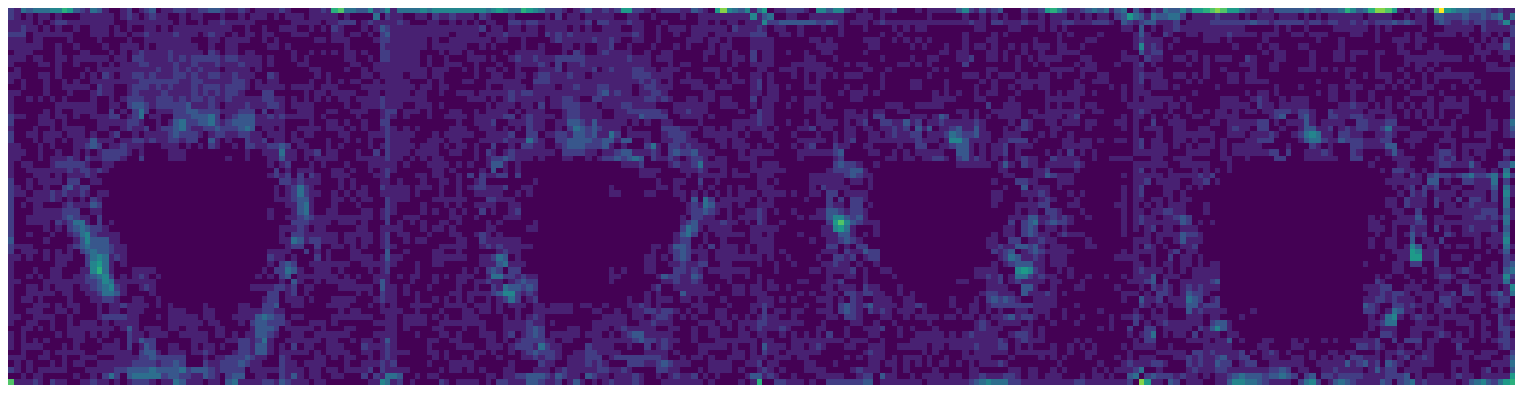}}

\caption{Visualisation of residuals between unvaccinated and vaccinated image samples.}
\label{fig:res}
\end{figure}

\subsubsection*{Vaccination}
Given a video footage or a still image, a preliminary procedure is to detect the face region (with any available face detection algorithms), crop a portrait (for each video frame) and resize it to meet the input resolution specification (e.g. $64 \times 64$ pixels), leaving aside the case of multiple faces per image for simplicity. The pre-processing step of vaccination prepares a mask by detecting facial landmarks with off-the-shelf face alignment algorithms and assigning binary digits to pixels inside/outside the face contour. The mid-processing step inputs both portrait and mask into the vaccinator and obtains a raw output. The post-processing step produces the vaccinated image by substituting the face region of the raw output with that of the original portrait according to the mask. It can be observed from Figure~\ref{fig:res} that vaccination introduces imperceptible perturbations to the non-face region.

\begin{figure}[t]
\centering
\begin{algorithm}[H]
\centering
\caption{Training (Cyber Immune System)}\label{alg:train}
\begin{algorithmic}

\Input $\boldsymbol{x}_{\circ} \in \mathcal{D}$
\Comment{image from dataset}
\Output $[\operatorname{Vaccinator}, \operatorname{Neutraliser}]$


\LineComment{vaccination}

\State $\boldsymbol{m} = \operatorname{MaskDetector}(\boldsymbol{x}_{\circ})$
\Comment{pre-proc.}
\State $\boldsymbol{x}_{\bullet}^{\text{raw}} = \operatorname{Vaccinator}(\boldsymbol{x}_{\circ}, \boldsymbol{m})$
\Comment{mid-proc.}
\State $\boldsymbol{x}_{\bullet} = \boldsymbol{x}_{\bullet}^{\text{raw}}\cdot \boldsymbol{m} + \boldsymbol{x}_{\circ} \cdot \bar{\boldsymbol{m}}$
\Comment{post-proc.}
\State $\mathcal{L}_{\text{imp}} = \operatorname{Distance}(\boldsymbol{x}_{\bullet}, \boldsymbol{x}_{\circ})$
\Comment{loss}
\State
\LineComment{neutralisation}
\State $\boldsymbol{m}^{\text{rnd}} = \operatorname{RandomAffine}(\boldsymbol{m})$
\Comment{pre-proc.}
\State $\boldsymbol{x}_{\bullet}^{\text{rnd}} = \operatorname{RandomTransform}(\boldsymbol{x}_{\bullet})$
\Comment{augmentation}
\LineComment{vaccinated case}
\State $\boldsymbol{y}_{\bullet}^{\text{raw}} = \operatorname{Neutraliser}(\boldsymbol{x}_{\bullet}^{\text{rnd}}, \boldsymbol{m}^{\text{rnd}})$
\Comment{mid-proc.}
\State $\boldsymbol{y}_{\bullet} = \boldsymbol{y}_{\bullet}^{\text{raw}}\cdot \boldsymbol{m}^{\text{rnd}} + \boldsymbol{x}_{\bullet} \cdot \bar{\boldsymbol{m}}^{\text{rnd}}$
\Comment{post-proc.}
\State $\mathcal{L}_{\text{rev}} = \operatorname{Distance}(\boldsymbol{y}_{\bullet}, \boldsymbol{x}_{\circ})$
\Comment{loss}
\LineComment{unvaccinated case}
\State $\boldsymbol{y}_{\circ}^{\text{raw}} = \operatorname{Neutraliser}(\boldsymbol{x}_{\circ}, \boldsymbol{m}^{\text{rnd}})$
\Comment{mid-proc.}
\State $\boldsymbol{y}_{\circ} = \boldsymbol{y}_{\circ}^{\text{raw}}\cdot \boldsymbol{m}^{\text{rnd}} + \boldsymbol{x}_{\circ} \cdot \bar{\boldsymbol{m}}^{\text{rnd}}$
\Comment{post-proc.}
\State $\mathcal{L}_{\text{val}} = \operatorname{Distance}(\boldsymbol{y}_{\circ}, \boldsymbol{x}_{\circ} \cdot \bar{\boldsymbol{m}}^{\text{rnd}})$
\Comment{loss}
\State
\LineComment{back-propagation}
\State $\mathcal{L} = \mathcal{L}_{\text{imp}} + \mathcal{L}_{\text{rev}} + \mathcal{L}_{\text{val}}$
\Comment{loss}
\State $\operatorname{Backprop}(\mathcal{L}, [\operatorname{Vaccinator}, \operatorname{Neutraliser}])$
\Comment{update}

\end{algorithmic}
\end{algorithm}
\hfill
\centering
\begin{algorithm}[H]
\centering
\caption{Inference (Vaccination)}\label{alg:inf_enc}
\begin{algorithmic}

\Input $\boldsymbol{x}_{\circ}$
\Comment{image (unvaccinated)}
\Output $\boldsymbol{x}_{\bullet}$
\Comment{image (vaccinated)}

\State $\boldsymbol{m} = \operatorname{MaskDetector}(\boldsymbol{x}_{\circ})$
\Comment{pre-proc.}
\State $\boldsymbol{x}_{\bullet}^{\text{raw}} = \operatorname{Vaccinator}(\boldsymbol{x}_{\circ}, \boldsymbol{m})$
\Comment{mid-proc.}
\State $\boldsymbol{x}_{\bullet} = \boldsymbol{x}_{\bullet}^{\text{raw}}\cdot \boldsymbol{m} + \boldsymbol{x}_{\circ} \cdot \bar{\boldsymbol{m}}$
\Comment{post-proc.}

\end{algorithmic}
\end{algorithm}
\hfill
\centering
\begin{algorithm}[H]
\centering
\caption{Inference (Neutralisation)}\label{alg:inf_dec}
\begin{algorithmic}

\Input $\boldsymbol{x}$
\Comment{image (attacked/not; vaccinated/not)}
\Output $\boldsymbol{y}$
\Comment{image (neutralised)}


\State $\boldsymbol{m} = \operatorname{MaskDetector}(\boldsymbol{x})$
\Comment{pre-proc.}
\State $\boldsymbol{y}^{\text{raw}} = \operatorname{Neutraliser}(\boldsymbol{x}\cdot \bar{\boldsymbol{m}}, \boldsymbol{m})$
\Comment{mid-proc.}
\State $\boldsymbol{y} = \boldsymbol{y}^{\text{raw}}\cdot \boldsymbol{m} + \boldsymbol{x} \cdot \bar{\boldsymbol{m}}$
\Comment{post-proc.}

\end{algorithmic}
\end{algorithm}
\end{figure}

\subsubsection*{Neutralisation}
The pre-processing step of neutralisation process prepares a mask and uses it to mask out the face region of the given image, which can be either vaccinated or unvaccinated. In the actual inference stage, the face region of the given image may be manipulated by arbitrary deepfake algorithms, which would cause slight misalignment of face landmarks in the neutralisation process. During the training stage, it is unnecessary to involve deepfake algorithms because such misalignment can be simulated by applying random affine transformations to the mask generated in vaccination process. We also apply random colour transformations to reinforce robustness against common image distortions. The mid-processing step inputs the masked image along with the mask into the neutraliser and yields a raw output. The post-processing step merges the face region of the raw output into the masked image, resulting in the neutralised image.

\begin{figure*}[t!] 
\centering
\subfloat[unvaccinated samples]{\includegraphics[width=1.98\columnwidth]{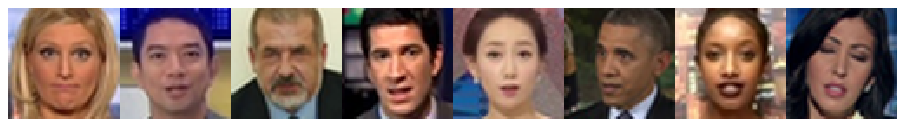}}
\\
\subfloat[neutralised samples with vaccination]{\includegraphics[width=1.98\columnwidth]{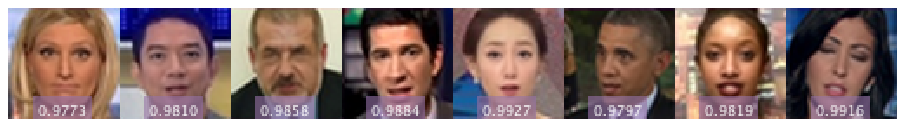}}
\\
\subfloat[neutralised samples without vaccination]{\includegraphics[width=1.98\columnwidth]{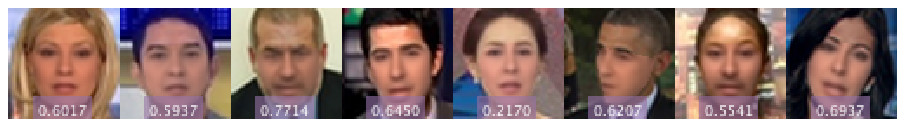}}

\caption{Comparison between neutralised image samples with and without vaccination. Numerical data denotes latent-space cosine similarity.}
\label{fig:wo_vacci}
\end{figure*}

\subsubsection*{Validation and Loss Functions}
Loss functions are essential for training neural networks. For cyber vaccination, we aim to achieve imperceptibility, reversibility and validatability. We evaluate imperceptibility by the similarity between the non-face region of the vaccinated image and that of the original images, and reversibility by the similarity between the face region of the neutralised image and that of the original images. There may be several feasible ways of imparting validatability to the system and a viable option is to let the neutraliser unresponsive to the unvaccinated input. In other words, the output is expected to remain a masked portrait when unvaccinated. It can then be identified effortlessly with the naked eye or automatically with a data-driven visual classifier. Let $\boldsymbol{x}_{\circ}$ and $\boldsymbol{x}_{\bullet}$ denote unvaccinated and vaccinated images, and $\boldsymbol{y}_{\circ}$ and $\boldsymbol{y}_{\bullet}$ their neutralised counterparts, respectively. Also let $\boldsymbol{m}$ denote a mask where the face region is assigned with one and non-face region with zero. Its inverse is denoted by $\bar{\boldsymbol{m}}$ where the binary assignment is opposite. Note that in practice the mask is not necessary to be composed of binary digits but of real numbers with soft edges for providing a smoother blending effect. The loss function for optimising both vaccinator and neutraliser is the (weighted) sum of three loss terms:
\begin{equation}
\mathcal{L} = \mathcal{L}_{\text{imp}}(\boldsymbol{x}_{\bullet}, \boldsymbol{x}_{\circ}) + \mathcal{L}_{\text{rev}}(\boldsymbol{y}_{\bullet}, \boldsymbol{x}_{\circ}) +\mathcal{L}_{\text{val}}(\boldsymbol{y}_{\circ}, \boldsymbol{x}_{\circ}\cdot\bar{\boldsymbol{m}}) .
\end{equation}
In our implementation, each loss term is composed of mean absolute error (MAE) and structural similarity index measure (SSIM). The algorithmic procedures for training and inference (vaccination and neutralisation) are shown in Algorithms~\ref{alg:train},~\ref{alg:inf_enc} and~\ref{alg:inf_dec}. For training an automatic validator to classify between vaccinated and unvaccinated images, we use neutralised images as the inputs and binary cross entropy as the loss function.

\section{Evaluation}
\label{sec:exp}
We evaluate the proposed cyber vaccination system with respect to imperceptibility, reversibility, robustness and validatability. We also carry out case studies for demonstrating deepfake immunity and discusses potential limitations.



\subsection{Experimental Setup}
The models are trained and tested on the FaceForensics++ dataset, consisting of a thousand video clips with trackable faces~\cite{9010912}. For both vaccinator and neutraliser, we use the U-Net as the backbone architecture~\cite{Ronneberger:2015aa} and apply a state-of-the-art version by OpenAI with residual connection and multi-head attention mechanisms~\cite{pmlr-v139-nichol21a}, which has been widely used as diffusion probabilistic models for various computer vision tasks~\cite{10.5555/3495724.3496298, 9878449, 9887996}. We specify the number of residual blocks to 3 and the attention resolutions to 4, 8 and 16. For the validator, we test serval representative classification models from pioneering to contemporary architectures, including multi-layer perceptron (MLP)~\cite{10.5555/104279.104293}, neural network by LeCun et al. (LeNet)~\cite{726791}, residual neural network (ResNet)~\cite{7780459}, vision transformer (ViT)~\cite{Dosovitskiy:2021aa} and next-generation convolutional neural network (ConvNeXT)~\cite{9879745}. We test the immunity to face replacement with both mask-dependent and mask-independent deepfakes. For the former, we use the original deepfake method and train several pairs of autoencoders for swapping between different pairs of identities. For the the latter, we employ a pre-trained identity-agnostic SimSwap model~\cite{10.1145/3394171.3413630}. We also test the immunity to face reenactment by using a pre-trained X2Face model~\cite{10.1007/978-3-030-01261-8_41}.


\begin{figure*}[!t]
\centering
\begin{minipage}{.5\textwidth}
\centerline{\includegraphics[width=0.98\columnwidth]{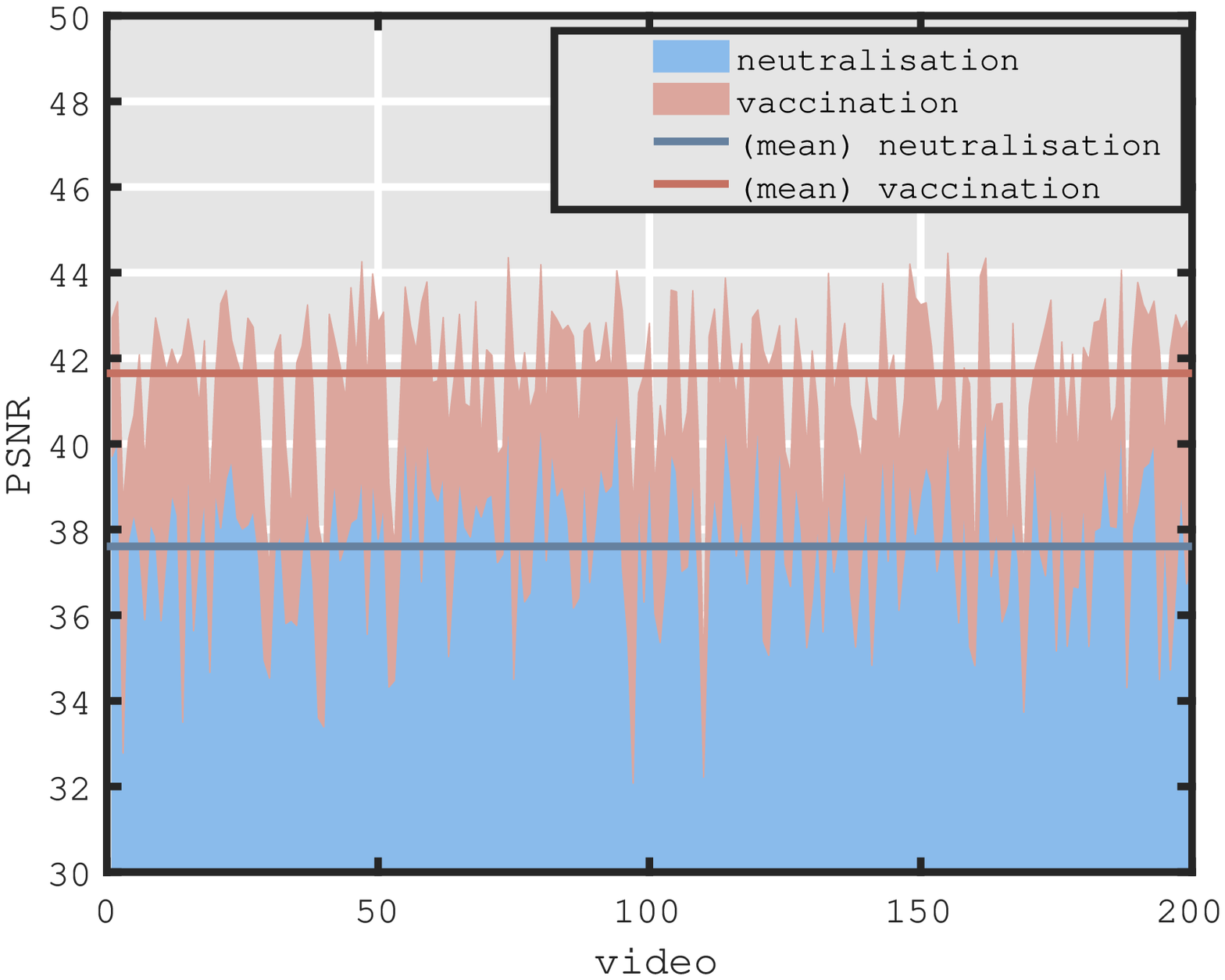}}
\caption{Evaluation of imperceptibility and reversibility.}
\label{fig:imp_rev}
\end{minipage}%
\begin{minipage}{0.5\textwidth}
\centerline{\includegraphics[width=0.98\columnwidth]{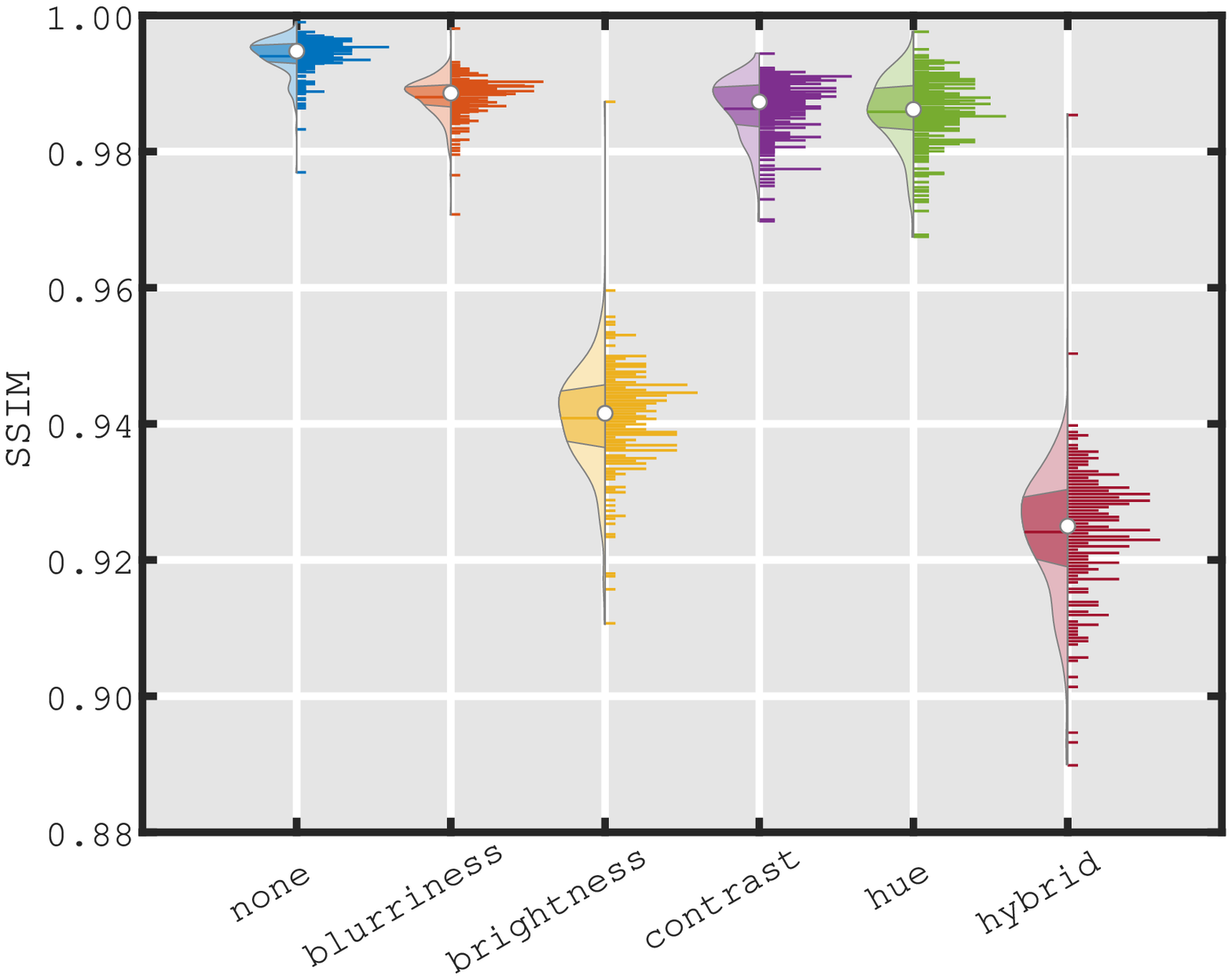}}
\caption{Evaluation of robustness with respect to different types of corruption.}
\label{fig:rob_all}
\end{minipage}
\end{figure*}




\subsection{Ablation Study on Cyber Vaccine}
We evaluate the effects from the cyber vaccine by making comparisons with a variant system without the vaccinator. In other words, this variant system only consists of a neutraliser trained for filling up the masked face area in such a way similar to image inpainting. For comparing between the neutralised results with and without vaccination, we measure the identity similarity. This is performed by projecting the images into a latent space using the FaceNet and computing the cosine similarity between the latent-space vectors~\cite{7298682}. It can be seen from Figure~\ref{fig:wo_vacci} that although the inpainting-based approach can reconstruct plausible portraits, the results are visually different from the original samples and the identity similarity is much lower than that from the vaccination-based approach. For 200 test samples, the average identity similarity with the vaccine is 0.99 and that without the vaccine is 0.57, suggesting that vaccination generally leads to a close identity similarity.

\subsection{Imperceptibility and Reversibility}
We require the vaccinator to keep distortion imperceptible and the neutraliser to reconstruct the original content with high fidelity. Imperceptibility and reversibility refer to the visual qualities of vaccinated images and neutralised images respectively in comparison to the original images. Peak signal-to-noise ratio (PSNR) is a common objective assessment of image quality. Figure~\ref{fig:imp_rev} presents the PSNR values of 200 video samples for which each value is averaged over all the frames. Typical PSNR values in lossy image/video compression are between 30 and 50 decibels (dB) under the condition of 8 bits per colour channel. The mean PSNR values of vaccinated and neutralised videos are above 40 and 35, respectively, indicating acceptable imperceptibility and reversibility.

\begin{figure*}[t!] 
\centering
\subfloat[blurriness]{\includegraphics[width=0.99\columnwidth]{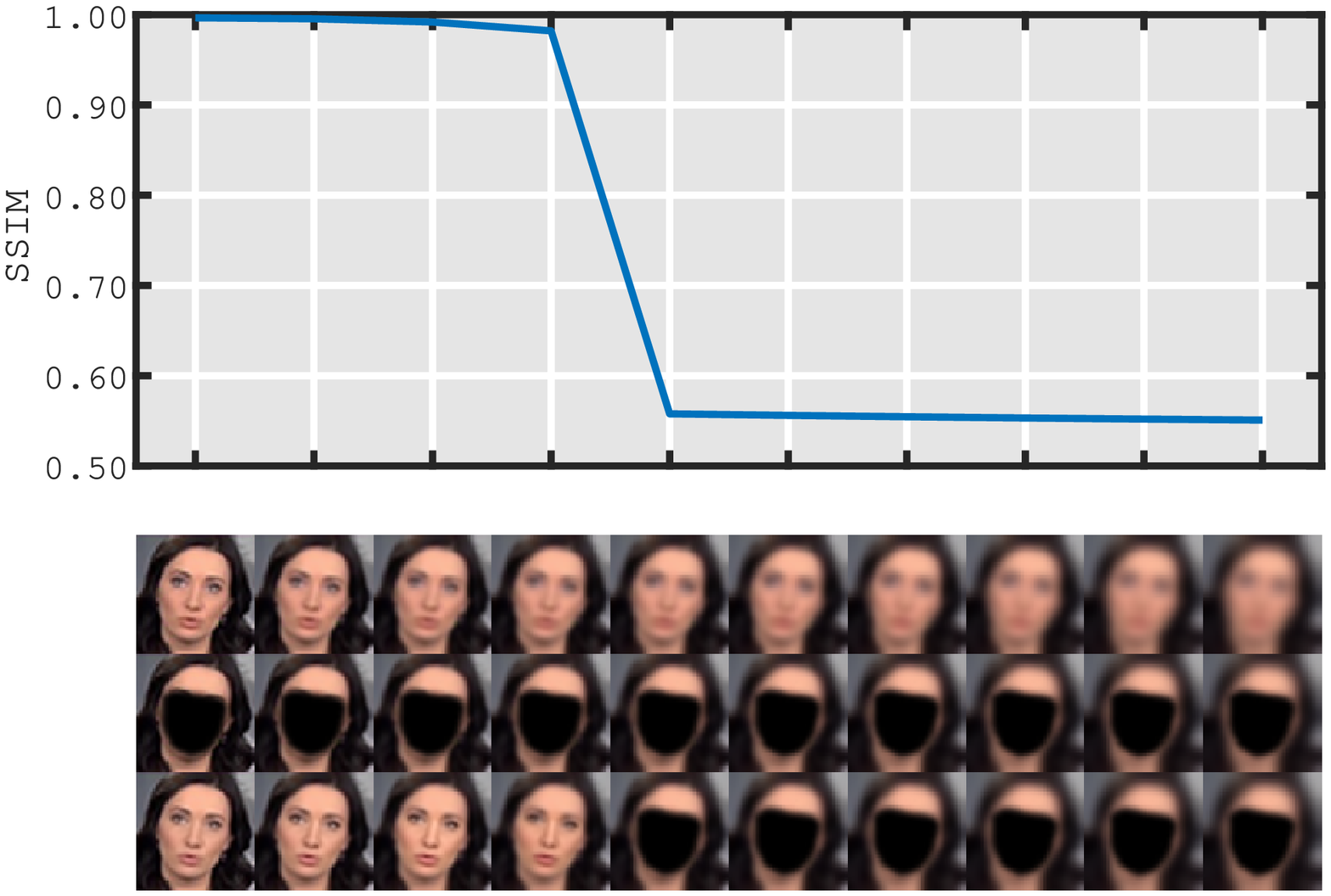}}
\hfil
\subfloat[brightness]{\includegraphics[width=0.99\columnwidth]{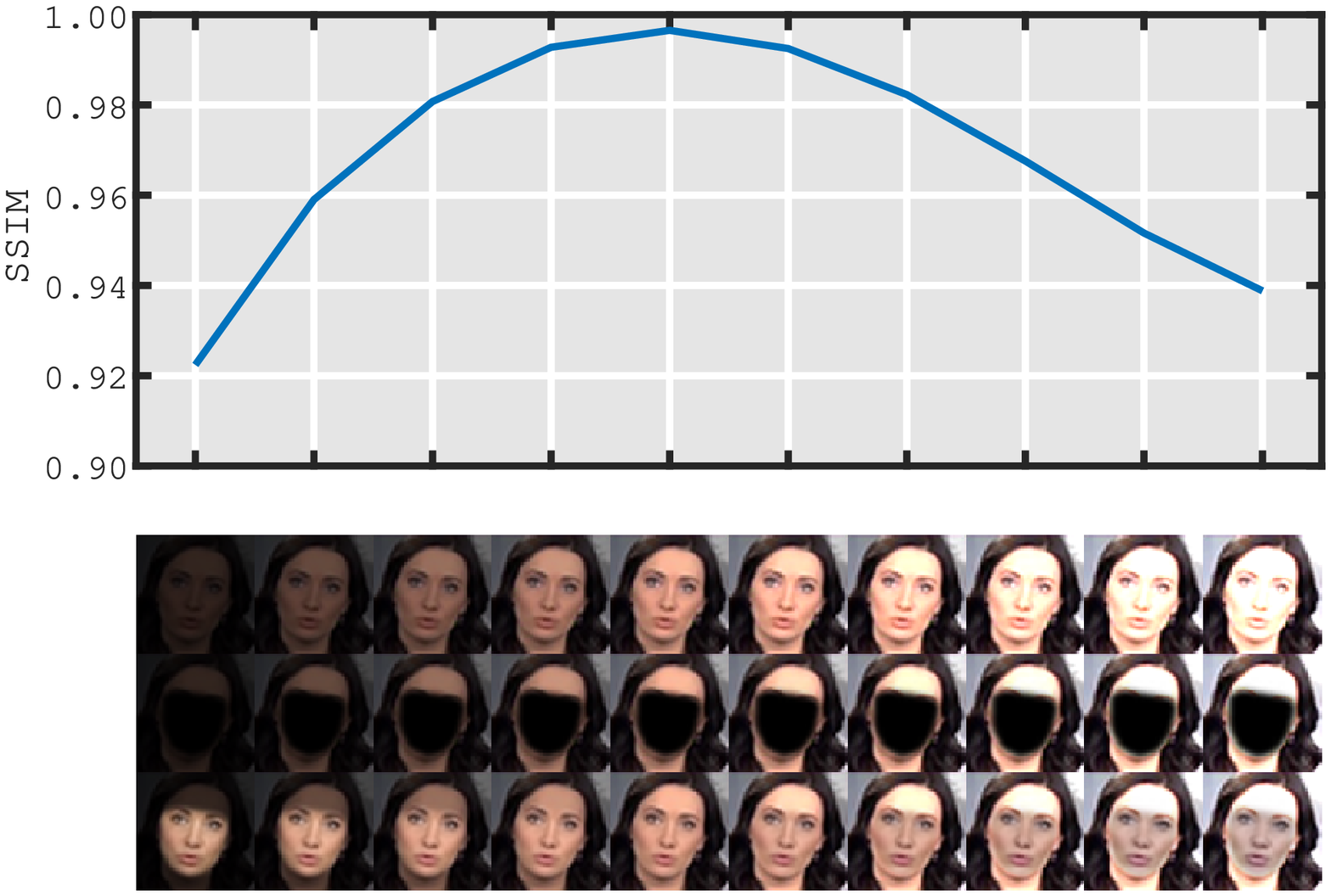}}
\\
\subfloat[contrast]{\includegraphics[width=0.99\columnwidth]{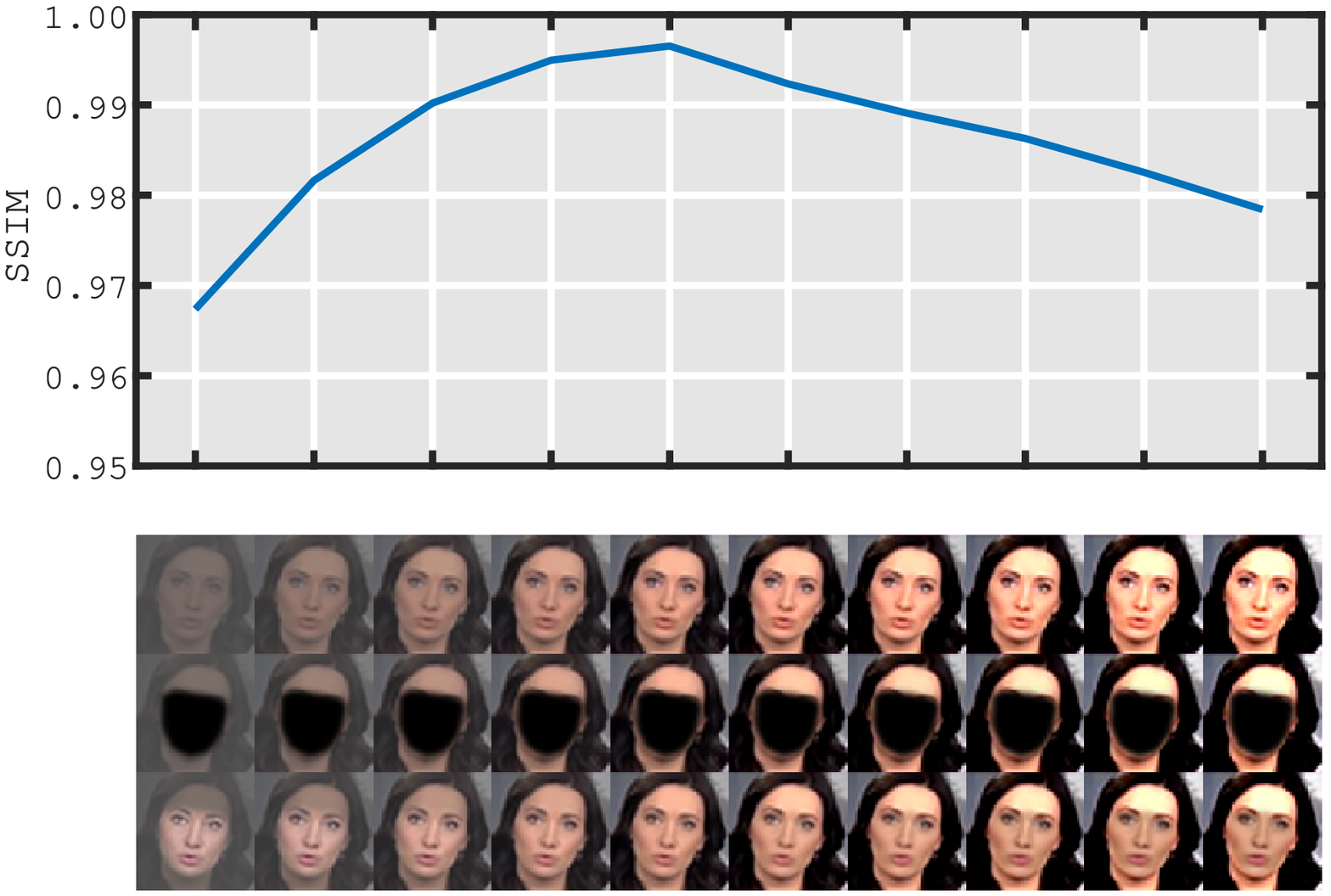}}
\hfil
\subfloat[hue]{\includegraphics[width=0.99\columnwidth]{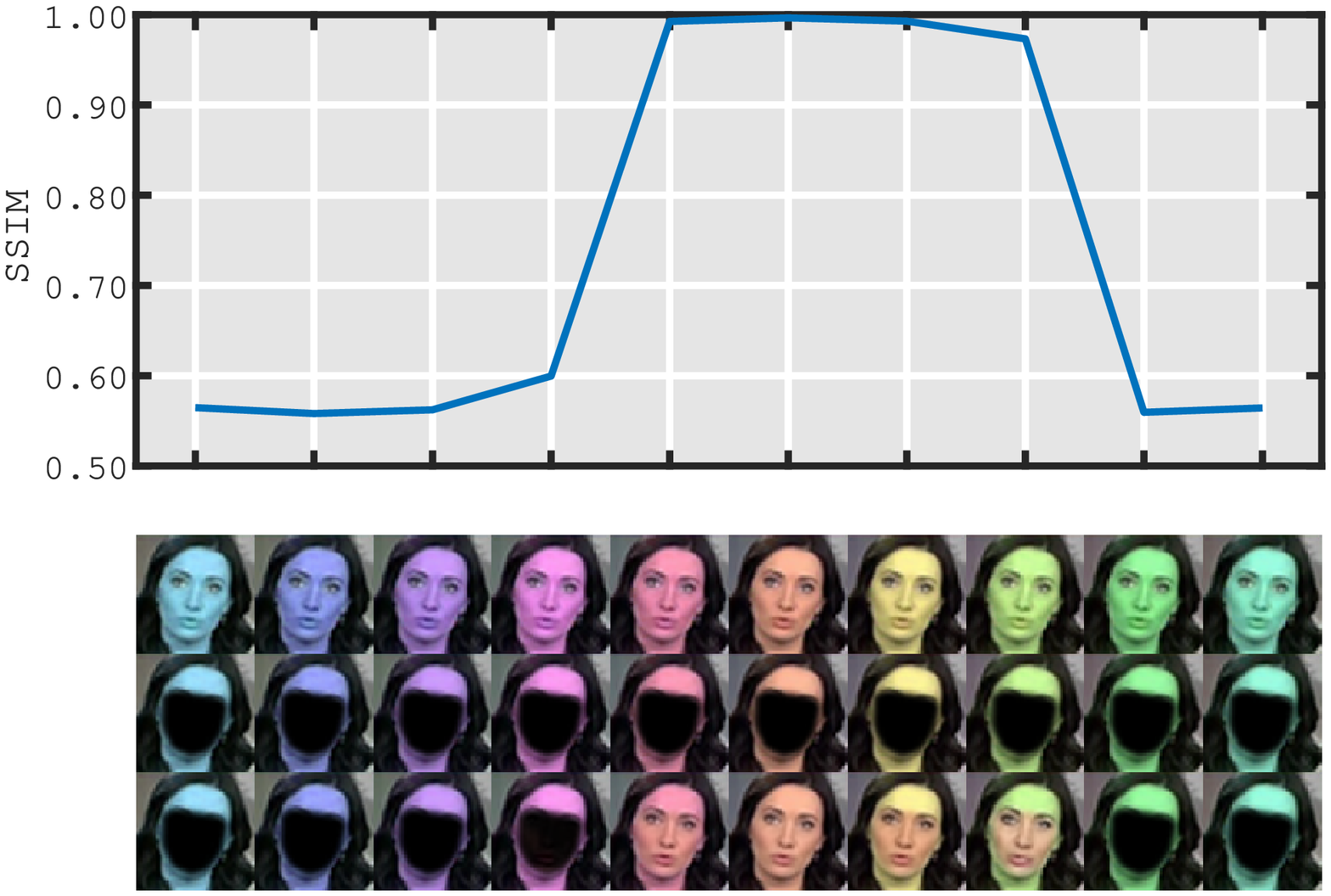}}

\caption{Evaluation of robustness with respect to full spectrum of degradation.}
\label{fig:rob_adjust}
\end{figure*}

\begin{figure*}[t!] 
\centering
\subfloat[none]{\includegraphics[width=0.66\columnwidth]{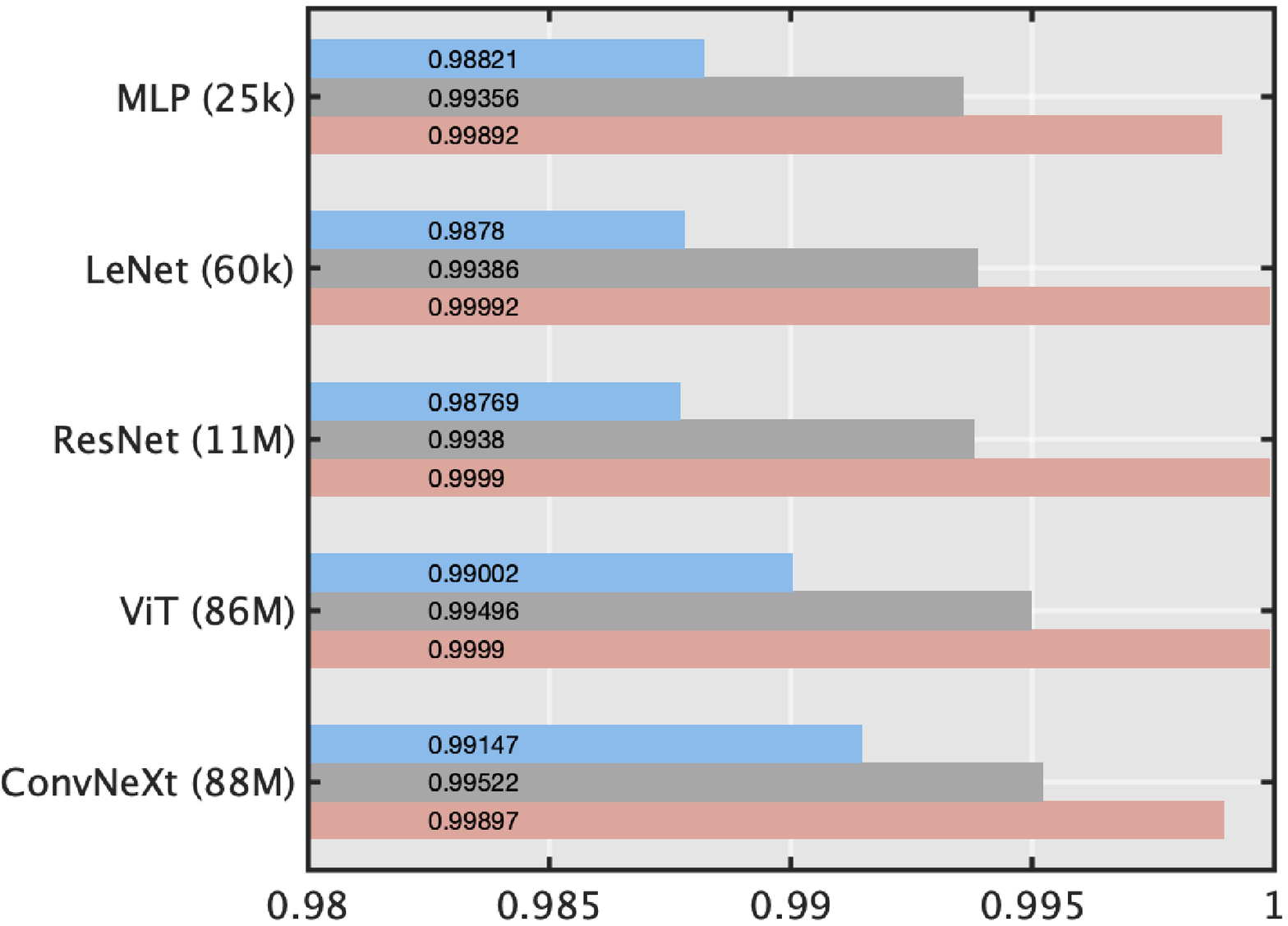}}
\hfil
\subfloat[blur]{\includegraphics[width=0.66\columnwidth]{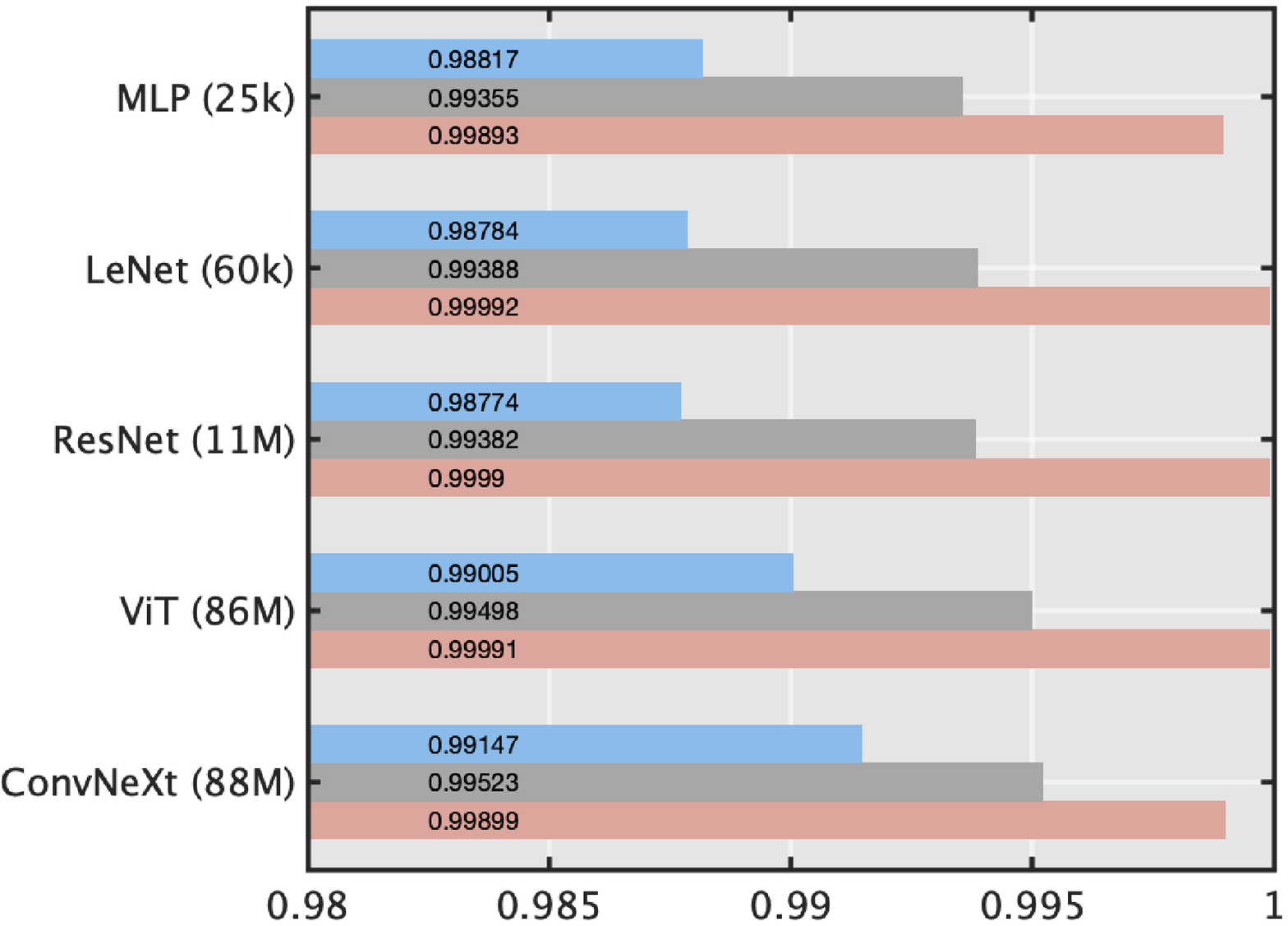}}
\hfil
\subfloat[brightness]{\includegraphics[width=0.66\columnwidth]{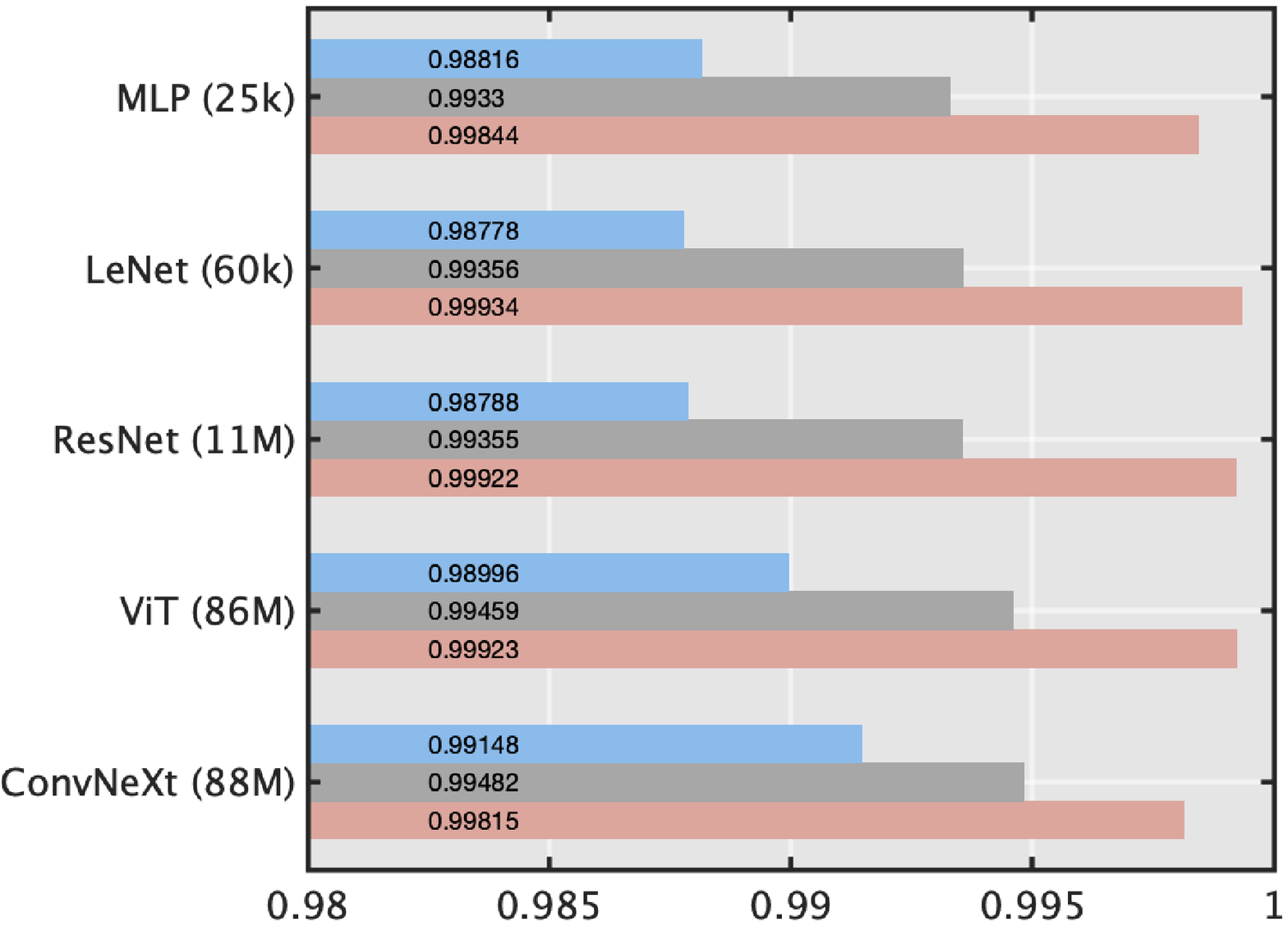}}
\\
\subfloat[contrast]{\includegraphics[width=0.66\columnwidth]{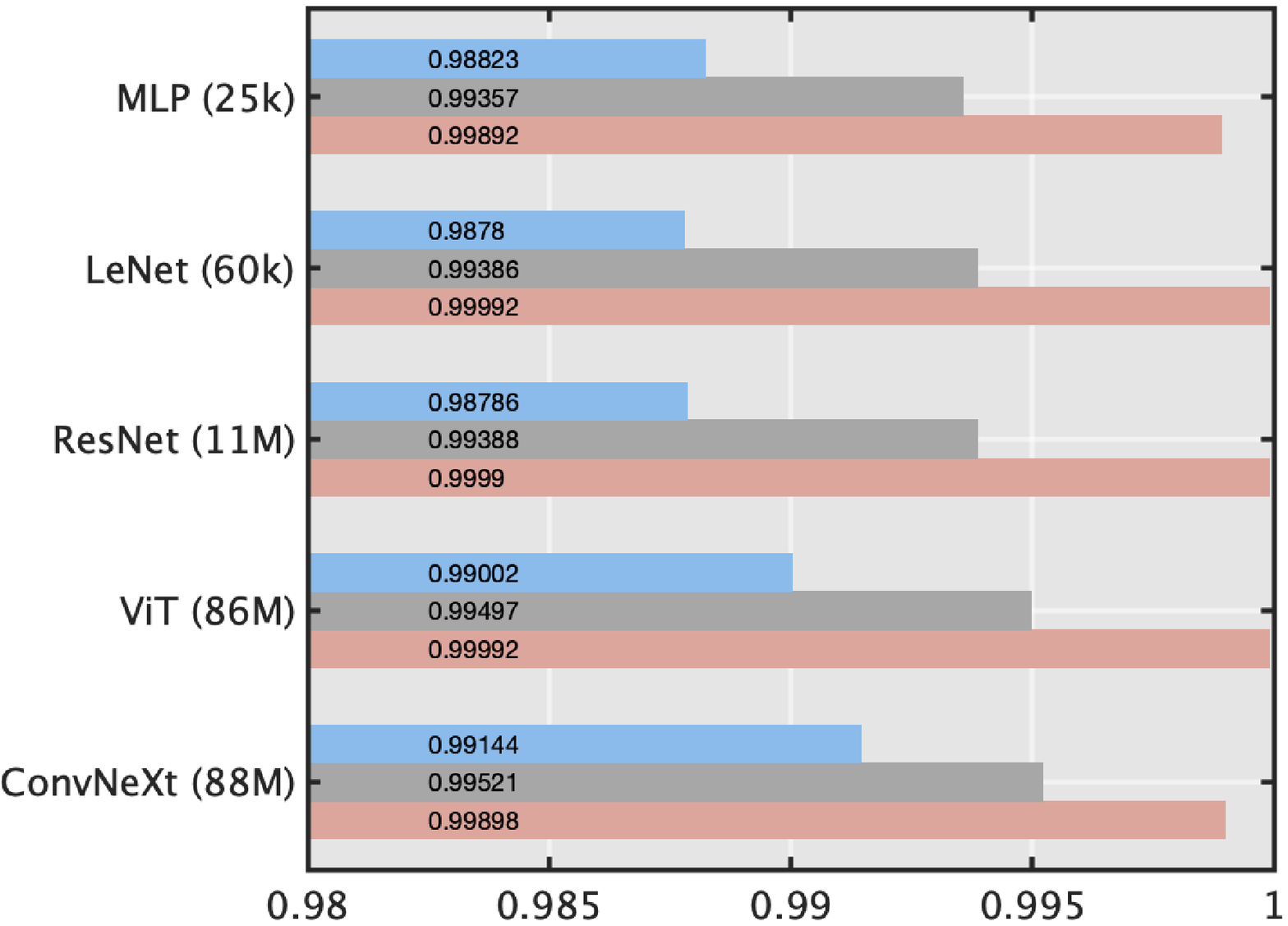}}
\hfil
\subfloat[hue]{\includegraphics[width=0.66\columnwidth]{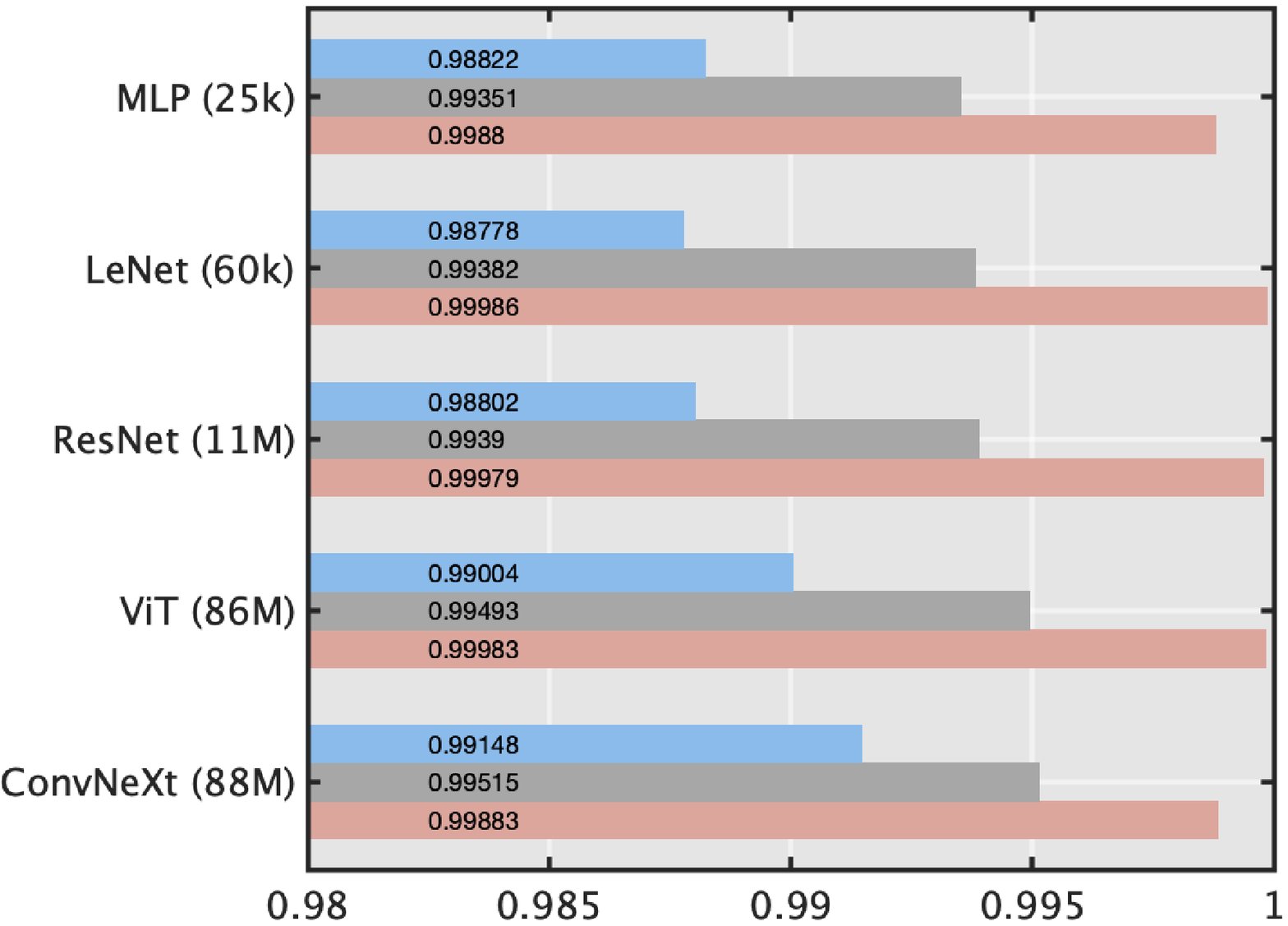}}
\hfil
\subfloat[hybrid]{\includegraphics[width=0.66\columnwidth]{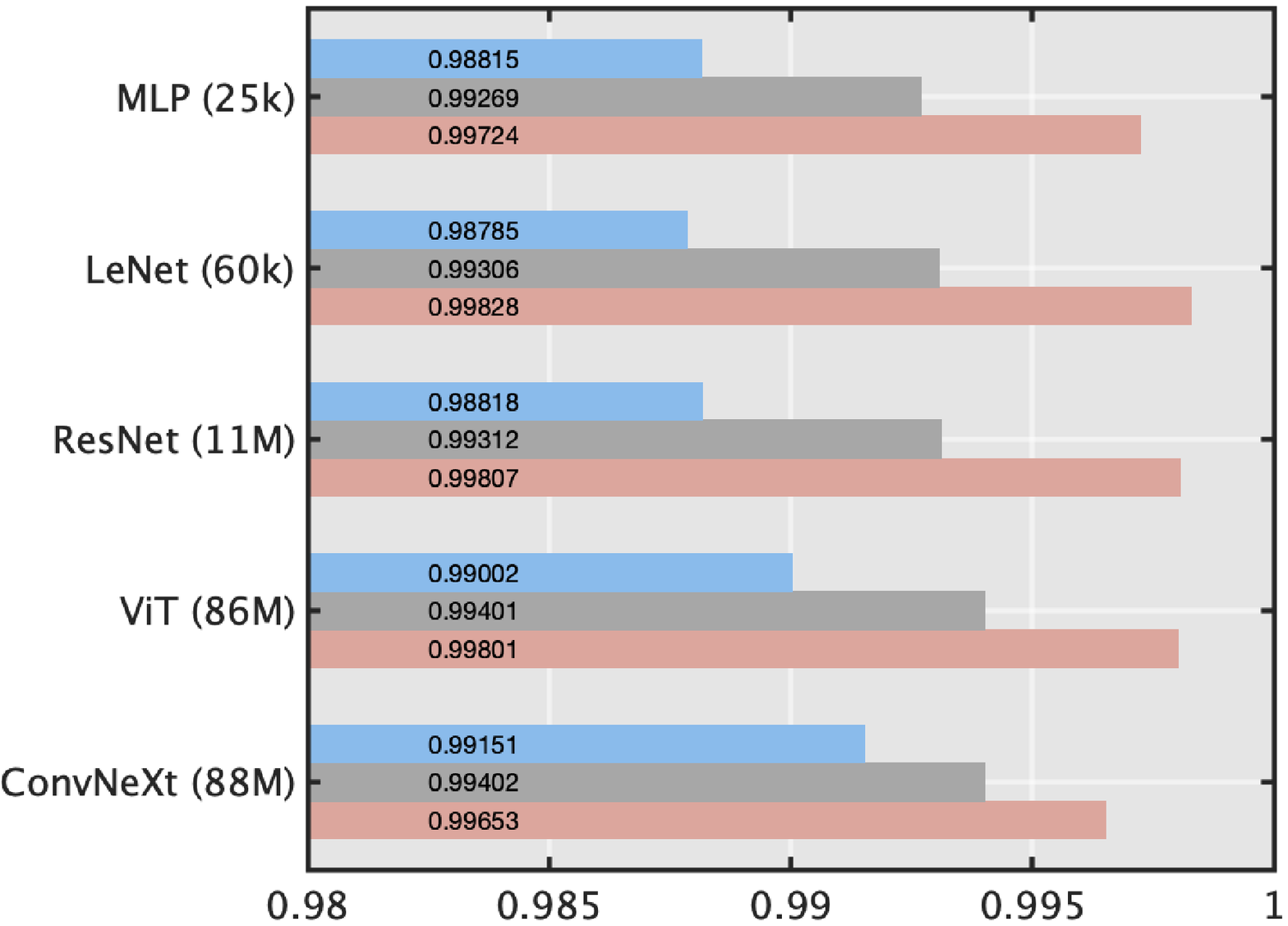}}

\caption{Validatability with different neural network classifiers in terms of true positive rate (top bar), accuracy (middle bar) and true negative rate (bottom bar).}
\label{fig:val_acc}
\end{figure*}

\begin{figure*}[t!] 
\centering
\subfloat[]{\includegraphics[width=0.66\columnwidth]{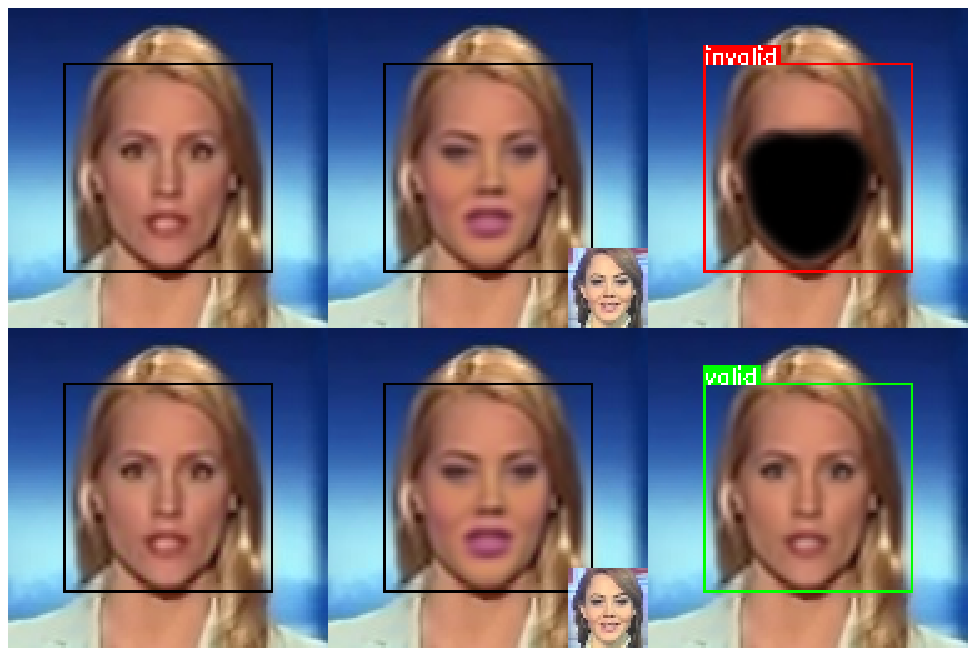}}
\hfil
\subfloat[]{\includegraphics[width=0.66\columnwidth]{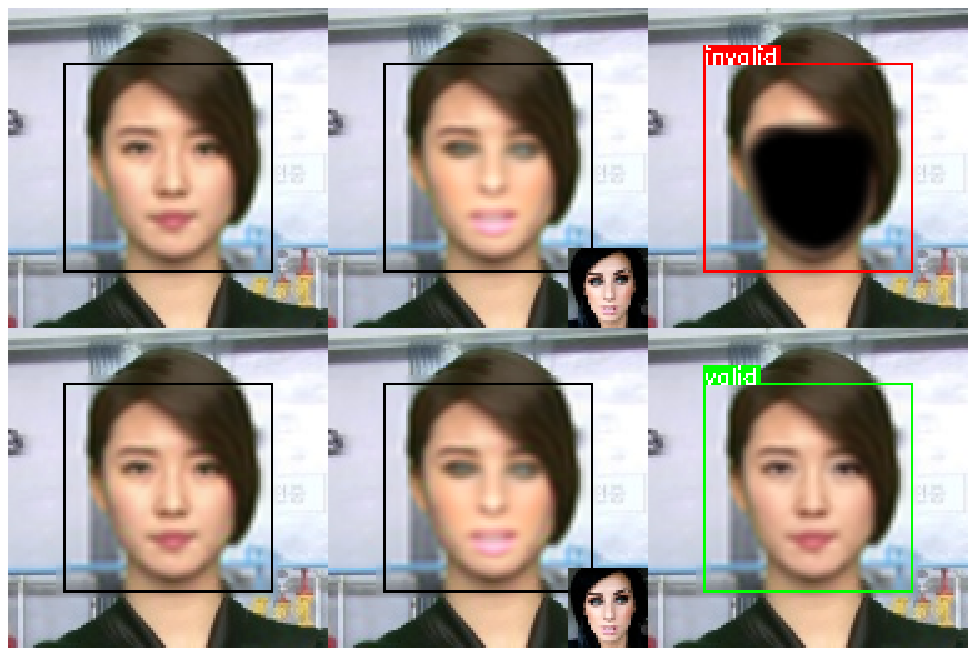}}
\hfil
\subfloat[]{\includegraphics[width=0.66\columnwidth]{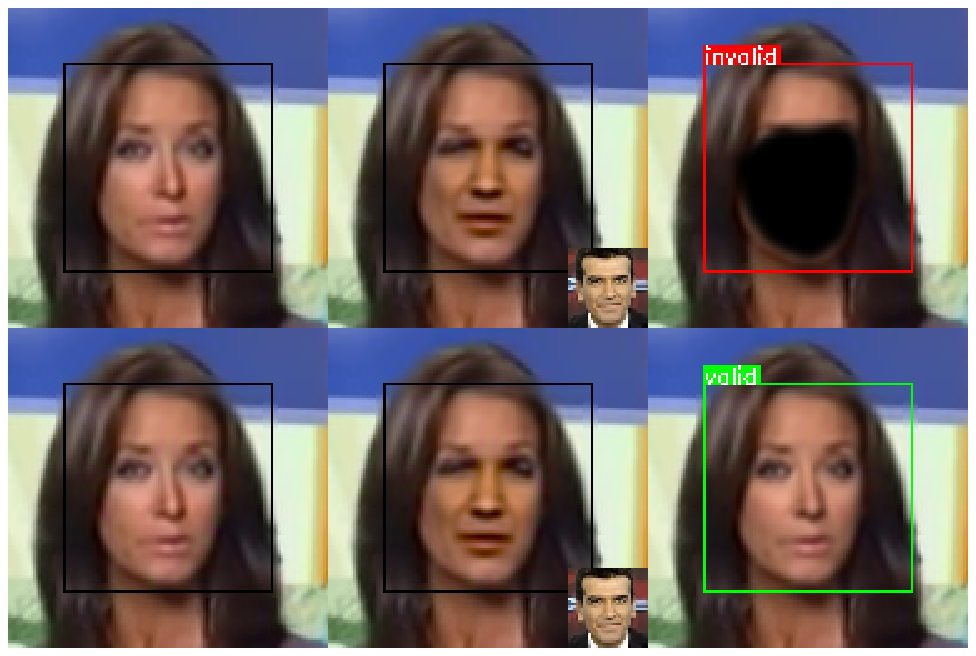}}
\\
\subfloat[]{\includegraphics[width=0.66\columnwidth]{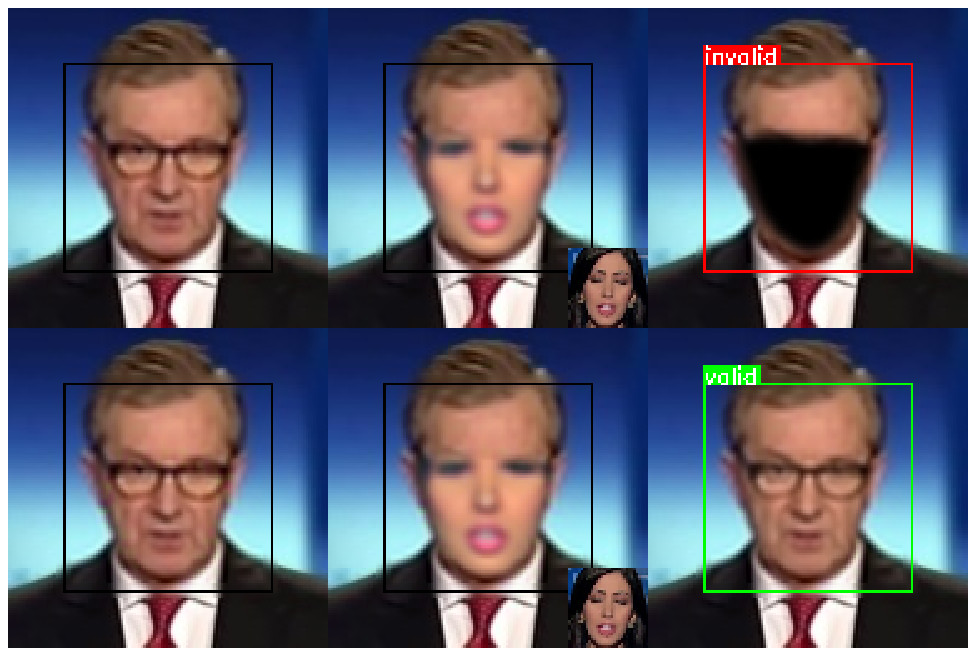}}
\hfil
\subfloat[]{\includegraphics[width=0.66\columnwidth]{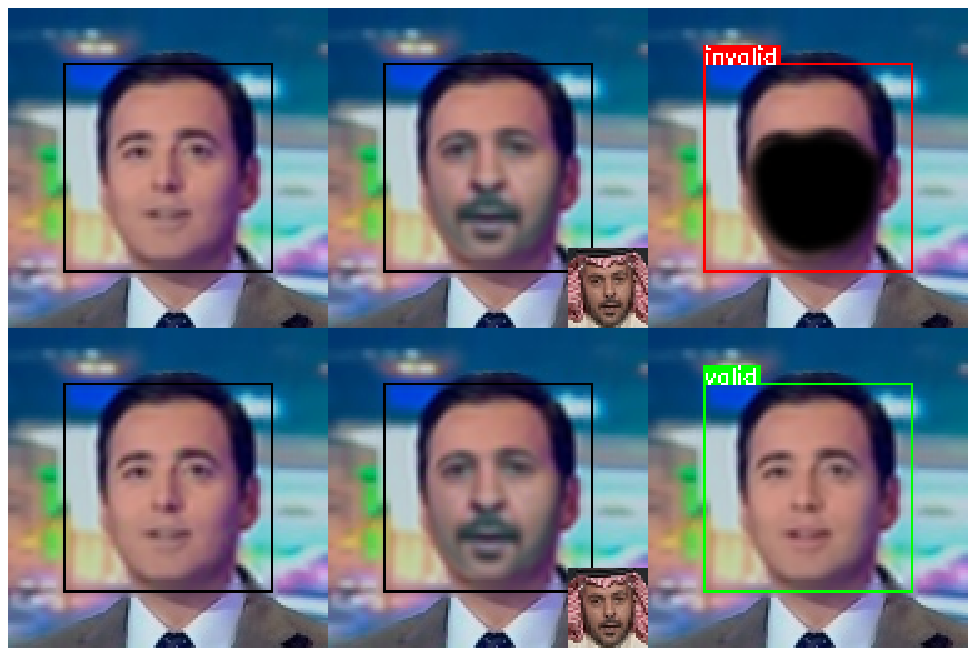}}
\hfil
\subfloat[]{\includegraphics[width=0.66\columnwidth]{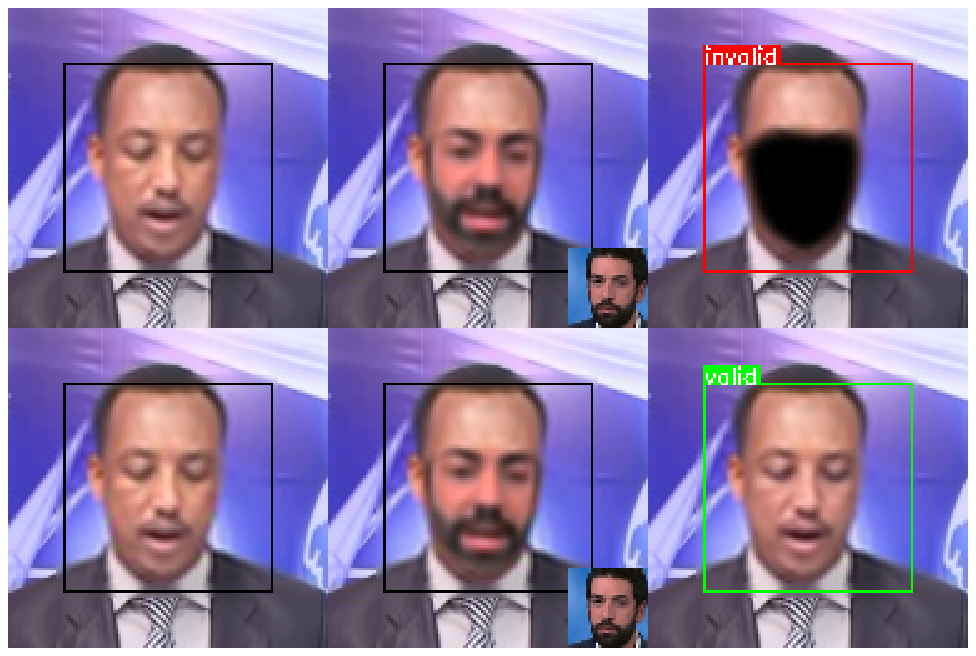}}

\caption{Demonstration of immunity to mask-dependent face replacement by the original deepfakes. Top (from left to right): unvaccinated videos, infected videos and neutralised videos. Bottom (from left to right): vaccinated videos, infected videos and neutralised videos.}
\label{fig:demo_immunity}
\end{figure*}

\begin{figure*}[t!] 
\centering
\subfloat[]{\includegraphics[width=0.5\columnwidth]{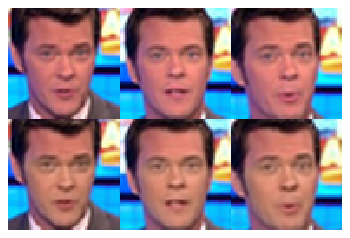}}
\hfil
\subfloat[]{\includegraphics[width=0.5\columnwidth]{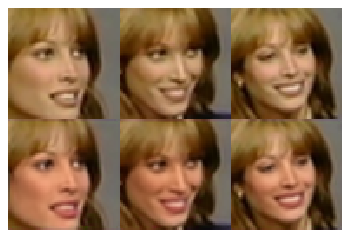}}
\hfil
\subfloat[]{\includegraphics[width=0.5\columnwidth]{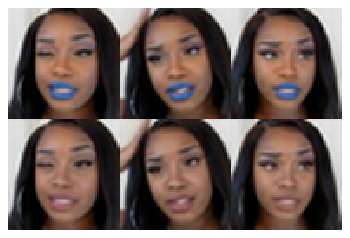}}
\hfil
\subfloat[]{\includegraphics[width=0.5\columnwidth]{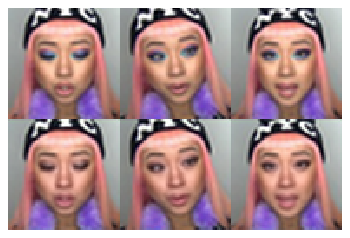}}
\hfil

\caption{Case study on colour misalignment between original videos (top row) and neutralised videos (bottom row).}
\label{fig:fail}
\end{figure*}

\subsection{Robustness}
Robustness is an important consideration when translating a system into practical applications. It concerns the extent to which a system can continue to function despite faults, disruptions and varying conditions. We measure the neutralisation performance by the SSIM of neutralised videos with regard to changes in blurriness, brightness, contrast and hue. Figure~\ref{fig:rob_all} compares each case in terms of the distribution of SSIM values. The degradation parameters are randomly sampled within a limited range. The average SSIM values for the cases of none and hybrid corruptions, as expected, are at opposite ends. Among individual cases, brightness adjustment appears to have the most negative effect to the neutralisation performance, whereas other adjustments cause minor fluctuations compared with the none corruption case. Figure~\ref{fig:rob_adjust} compares each case over the full spectrum of degradation. It can be seen that the system is capable of withstanding major changes in brightness and contrast, while sudden performance drops are observed in the presence of extreme adjustments of blurriness and hue. Overall, the evaluations show that the system can be considered reasonably resilient within a certain range of degradation and reliable for certain types of corruption.

\subsection{Validatability}
To demonstrate that the vaccinated videos can be automatically validated and readily distinguished from the unvaccinated ones, we apply several neural network classifiers with a wide variety of model size and architectural complexity and evaluate their classification performance across different degradation conditions. As shown in Figure~\ref{fig:val_acc}, the average accuracy of each classifier is all around 99 percent for every condition and the true positive rate is only slightly lower than the true negative rate. In general, it appears that an advanced model with a larger number of parameters and a more complex connection of neurones tends to achieve higher accuracy. Nevertheless, it is remarkable that even a most rudimentary perceptron model can demonstrate classification performance comparable to state-of-the-art models, suggesting that the neutralised results from vaccinated and unvaccinated videos are readily distinguishable.


\subsection{Deepfake Immunity and Limitations}
\begin{figure}[!t]
\centering
\centerline{\includegraphics[width=0.99\columnwidth]{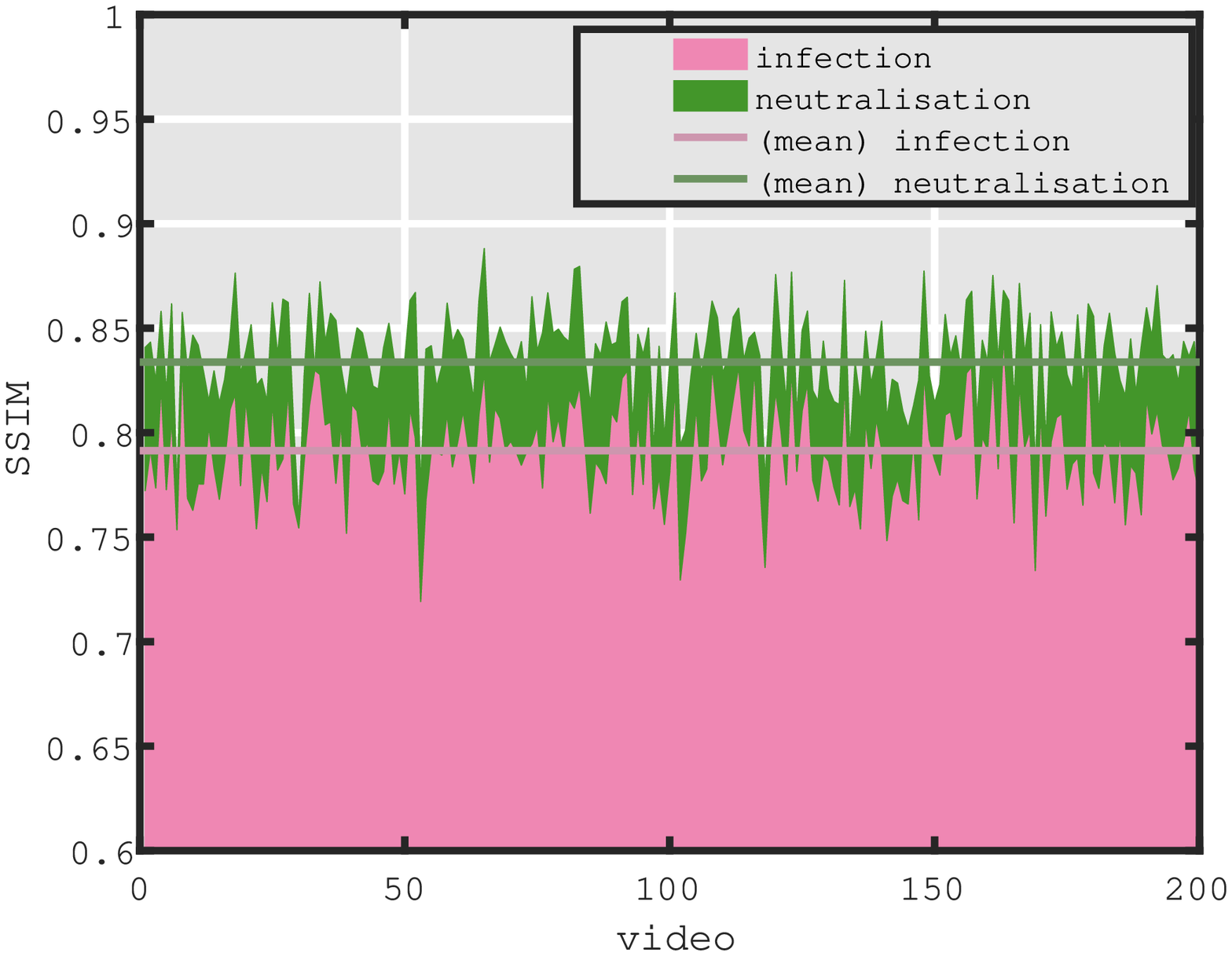}}
\caption{Evaluation of reversibility of infections by SimSwap.}
\label{fig:simswap_ssim}
\end{figure}

\begin{figure}[!t]
\centering
\centerline{\includegraphics[width=0.99\columnwidth]{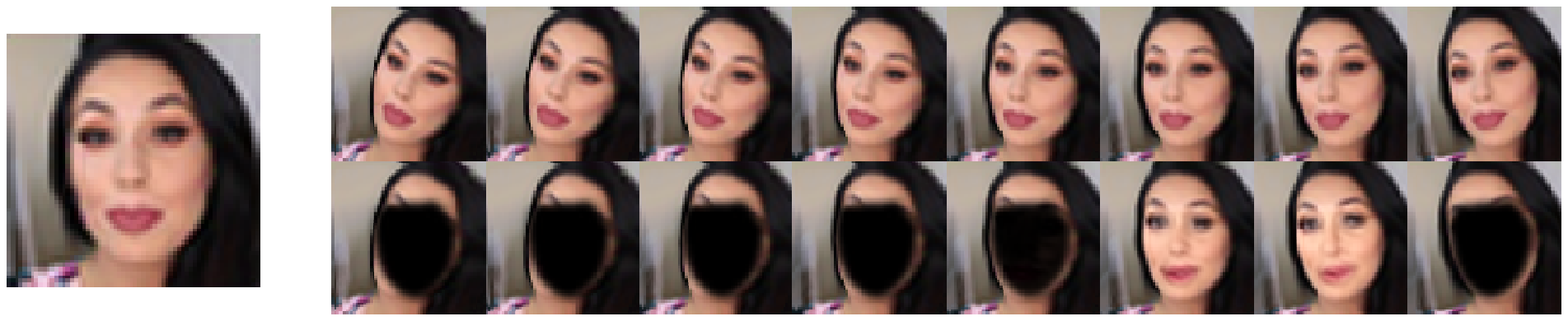}}
\caption{Demonstration of immunity to face reenactment by X2Face. Left: vaccinated image. Right: infected images (top row) and neutralised images (bottom row).}
\label{fig:x2face}
\end{figure}

We demonstrate immunity to face replacement and face reenactment, which are the major types of deepfakes that could lead to treacherous consequences. We evaluate the immunity to both mask-dependent and mask-independent deepfake methods. In general, mask-dependent deepfakes would be easier to cope with since the non-face region is more likely to be kept intact. For mask-dependent face replacement, we employ the classic autoencoder-based method. From Figure~\ref{fig:demo_immunity}, we can see that the facial area is restored with high fidelity, providing that the videos are vaccinated prior to attacks. By contrast, an empty facial area is observed for the unvaccinated cases, suggesting that the system shows no immune response when the given videos are unvaccinated. The bounding box annotations on the neutralised frames show the validity of vaccination determined by a validator model. It is worth noting that the system also shows adaptability to head positions and occlusion objects (e.g. eyeglasses). There are nonetheless some limitations such as inaccurate skin tones and mismatched make-up colours, as shown in Figure~\ref{fig:fail}. The former problem may be improved through post-processing by using a more delicate blending mechanism. The latter phenomenon is likely due to unusual colours, namely out-of-distribution data, and hence a possible improvement is to train the models with a greater diversity of data collected from real source or created artificially. For mask-independent face replacement, we use a pre-trained SimSwap model, which is an identity-agnostic model, being able to swap arbitrary identities with a single model rather than requiring one model for each pair of identities. Figure~\ref{fig:simswap_ssim} shows an increase of SSIM scores between the infected and neutralised videos. From the visual examples provided in Figure~\ref{fig:simswap_ssim}, it can be seen that while the neutralisation is functioning, the reversibility in the mask-independent case is inferior to that in the mask-dependent case, especially in the boundary of face area (e.g. eyebrows). We also test the performance against face reenactment with a pre-trained X2Face model. In particular, it is interesting to explore the impact of pose changes. The results from Figure~\ref{fig:x2face} suggests that neutralisation would be viable in the case of minor pose changes and yet infections are irreversible when major pose changes are presented.



\section{Conclusion}
\label{sec:con}
In this study, we propose a cyber vaccination mechanism for conferring immunity to deepfakes. It is shown that the distortion caused by vaccination is generally imperceptible and effective neutralisation is achieved under various corruption conditions. Furthermore, the validity of vaccination can be readily verified with a wide range of neural network classifiers. There is nonetheless colour misalignment in certain cases, which may be improved through post-processing and data augmentation. Despite the dependency on masks, the immunity to mask-independent face replacement is demonstrated. For face reenactment that causes changes in pose and thus major alterations in non-face region, novel mechanisms with greater robustness are worth further investigation. We envisage further progress in cyber vaccines for addressing more in-the-wild threats imposed by deepfakes.

\begin{figure}[!t]
\centering
\centerline{\includegraphics[width=0.99\columnwidth]{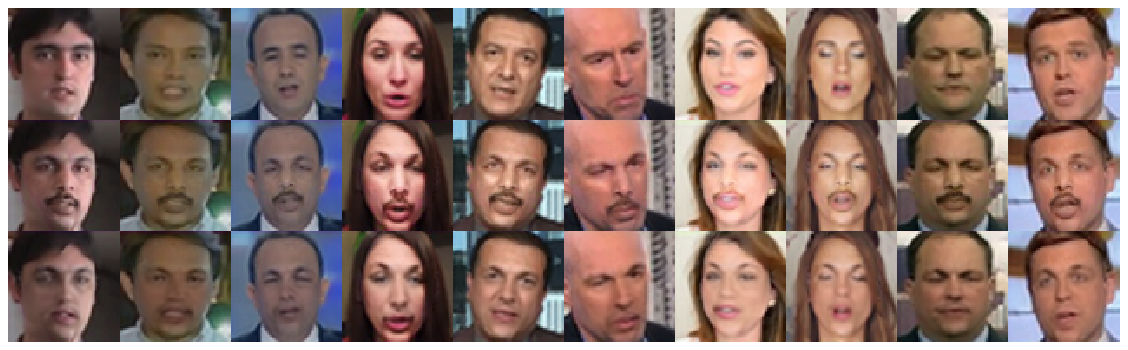}}
\caption{Demonstration of immunity to mask-independent face replacement by SwimSwap. Top: original images. Middle: infected images. Bottom: neutralised images.}
\label{fig:simswap_montage}
\end{figure}

\bibliographystyle{Transactions-Bibliography/IEEEtran}
\bibliography{./Bib/deepfake_bib}

\end{document}